\documentclass[final]{siamltex}

\usepackage{epsf}
\usepackage{amssymb}
\usepackage{amsbsy}
\usepackage{verbatim}

\usepackage{graphicx}

\usepackage{datetime}

\usepackage[colorlinks=true, linkcolor=blue, citecolor=blue, filecolor=blue, runcolor=blue, urlcolor = blue]{hyperref}

\newcommand{\OpFldDissp}{\mathcal{L}}
\newcommand{\OpFSDissp}{\Upsilon}
\newcommand{\OpFSCouple}{\Gamma}
\newcommand{\OpSFCouple}{\Lambda}

\newcommand{\OpFldDisspDiscr}{L}
\newcommand{\OpFSDisspDiscr}{\Upsilon}
\newcommand{\OpFSCoupleDiscr}{\Gamma}
\newcommand{\OpSFCoupleDiscr}{\Lambda}

\newcommand{\OpAdjoint}{{\dagger}}
\newcommand{\RegAdjoint}{{T}}

\newcommand{\Eng}{\Phi}

\newcommand{\mb}[1]{\mathbf{#1}}

\newcommand{\cm}{\mbox{\tiny cm}}
\newcommand{\Xcm}{\mbox{$\mathbf{X}_{\cm}$}}
\newcommand{\Ang}{\mbox{$\mathbf{\Theta}$}}

\newcommand{\subtxt}[1]{ {\mbox{\tiny #1}} }

\ifpdf
\fi

\begin{document}

\title{Stochastic Eulerian Lagrangian Methods for Fluid-Structure Interactions \\ with Thermal Fluctuations}

\author{Paul J. Atzberger
\thanks{University of California, 
Department of Mathematics , Santa Barbara, CA 93106; 
e-mail: atzberg@math.ucsb.edu; phone: 805-893-3239;
Work supported by NSF CAREER Grant DMS - 0956210.
\href{http://www.math.ucsb.edu/\~atzberg/}{
$\mbox{http://www.math.ucsb.edu/$\sim$atzberg/}$
}
}
}

\maketitle

\begin{abstract}
We present approaches for the study of fluid-structure interactions 
subject to thermal fluctuations.  A mixed mechanical description
is utilized combining Eulerian and Lagrangian reference frames.
We establish general conditions for operators coupling these descriptions.
Stochastic driving fields for the formalism 
are derived using principles from statistical mechanics.  The
stochastic differential equations of the formalism are found 
to exhibit significant stiffness in some physical regimes.  
To cope with this issue, we derive reduced stochastic differential 
equations for several physical regimes.  
We also present stochastic numerical methods for each regime to approximate 
the fluid-structure dynamics and to generate efficiently the required stochastic 
driving fields.  To validate the methodology in each regime, we perform analysis 
of the invariant probability distribution of the stochastic dynamics of 
the fluid-structure formalism.  We compare this analysis with results from 
statistical mechanics.  To further demonstrate the applicability of the methodology, we perform
computational studies for spherical particles having translational and 
rotational degrees of freedom.  We compare these studies with results from 
fluid mechanics.  The presented approach provides for fluid-structure 
systems a set of rather general computational methods for treating 
consistently structure mechanics, hydrodynamic coupling, and 
thermal fluctuations.  
\end{abstract}

\begin{keywords}
Fluid-Structure Interaction, Statistical Mechanics, Fluid Dynamics, Thermal Fluctuations,
Fluctuating Hydrodynamics, Stochastic Eulerian Lagrangian Method, SELM.
\end{keywords}


\pagestyle{myheadings}
\thispagestyle{plain}
\markboth{P.J. ATZBERGER}
{STOCHASTIC EULERIAN LAGRANGIAN METHODS}

\section{Introduction} 

The development of analytic and computational approaches for the 
study of fluid-structure interactions has a rich history.
Motivations for past work in this area include the study of aerodynamic 
oscillations induced in airplane wings and propellers~\cite{Dowell2001, Fung1955}, 
the study of animal locomotion including swimming and insect
flight~\cite{Miller2005,Lauga2009,Purcell1977}, and the study of physiological problems 
such as blood flow through heart valves~\cite{Griffith2009, Peskin1977, Givelberg2003}. 
A central challenge in work on these applications has been to develop descriptions 
which capture essential features of the fluid structure interactions while 
introducing approximations which facilitate analysis and the 
development of tractable numerical methods~\cite{Dowell2001, Mittal2005}. 
Many such challenges remain and this area of research is still very 
active~\cite{Griffith2009, Ceniceros2009, Mittal2005, Braescu2007}.  Recent 
scientific and technological advances motivate the study of fluid-structure
interactions in new physical regimes often involving very small length 
scales~\cite{Squires2005, Watari2008, Crocker2000, Moffitt2008}.  
At sufficiently small length scales thermal fluctuations play an important 
role and pose additional challenges in the study of fluid-structure systems.

Significant past work has been done on the formulation of descriptions
of fluid-structure interactions subject to thermal fluctuations.  
Many of these analytic and numerical approaches originate from 
the polymer physics community~\cite{Doi1986, Ermak1978, Bird1987, Brady1988}.  
To obtain descriptions
tractable for analysis and numerical simulation, these approaches typically 
place an emphasis on approximations which retain only the structure 
degrees of freedom.  This often results in significant simplifications in the 
descriptions and in significant computational savings.  This eliminates the many 
degrees of freedom associated with the fluid and avoids having to resolve the potentially 
intricate and stiff stochastic dynamics 
of the fluid.  These approaches have worked especially well for the study of 
bulk phenomena in free solution and the study of complex fluids and soft 
materials~\cite{Bird1987, Doi1986, Larson1999}.  

Recent applications arising in the sciences and in technological fields
present situations in which resolving the dynamics of the fluid may be
important and even advantageous both for modeling and computation.  
This includes modeling the spectroscopic responses of biological 
materials~\cite{Brown2010NSE, Gotter1996, Mezei2003},
studying transport in microfluidic and nanofluidic devices~\cite{Squires2005, Pennathur2010},
and investigating dynamics in biological systems~\cite{Alberts2002, Danuser2006}.
There are also other motivations for representing the fluid explicitly
and resolving its stochastic dynamics.  This includes the development of hybrid 
fluid-particle models in which thermal fluctuations 
mediate important effects when coupling continuum and 
particle descriptions~\cite{DeFabritiis2007, Donev2010}, the study of 
hydrodynamic coupling and diffusion in the vicinity of surfaces 
having complicated geometries~\cite{Squires2005}, 
and the study of systems in which there are many 
interacting mechanical structures~\cite{Banchio2003,Peskin2002,Peskin1977}.  
To facilitate the development of methods for studying such phenomena in 
fluid-structure systems, we present a rather general formalism which 
captures essential features of the coupled stochastic dynamics of the fluid and 
structures.

To model the fluid-structure system, a mechanical description
is utilized involving both Eulerian and Lagrangian reference 
frames.  Such mixed descriptions arise rather naturally, since 
it is often convenient to describe the structure configurations in a 
Lagrangian reference frame while it is convenient to describe
the fluid in an Eulerian reference frame.  In practice, this   
presents a number of challenges for analysis and numerical 
studies.  A central issue concerns how to couple the descriptions
to represent accurately the fluid-structure interactions,
while obtaining a coupled description which can be treated 
efficiently by numerical methods.
Another important issue concerns how to account properly for 
thermal fluctuations in such approximate descriptions.  This must
be done carefully to be consistent with statistical mechanics. 
A third issue concerns the development of efficient computational methods.  This requires discretizations of 
the stochastic differential equations and the development of efficient 
methods for numerical integration and stochastic field generation.

We present a set of approaches to address these 
issues.  The formalism and general
conditions for the operators which couple the 
Eulerian and Lagrangian descriptions are presented 
in Section~\ref{sec_summary_SELM}.  We discuss simplified descriptions of
the fluid-structure system in different physical regimes in Section~\ref{sec_derivations}.
A derivation of the stochastic driving fields used to represent 
the thermal fluctuations is also presented in Section~\ref{sec_derivations}.  Stochastic
numerical methods are discussed for the approximation of the stochastic dynamics 
and generation of stochastic fields in Sections~\ref{sec_comp_method}.
To validate the methodology, we perform in each regime analysis of 
the invariant probability distribution of the stochastic dynamics of 
the fluid-structure formalism.  
We compare this analysis with results from statistical mechanics in 
Section~\ref{sec_equil_stat_mech}.  To demonstrate the applicability of 
the methodology, we perform computational studies for spherical particles 
having translational and rotational degrees of freedom.  We compare these 
computational studies with results from fluid mechanics in 
Section~\ref{sec_applications}.

It should be mentioned that related computational methods have been 
introduced for the study of fluid-structure interactions
~\cite{Peskin2002, Arienti2003, Bodard2006, Braescu2007, 
Wang2009, Mittal2005, Zohdi2005, Kim2007, Lim2008}.
In recent papers, significant work also has been done toward 
incorporating the role of thermal 
fluctuations~\cite{Atzberger2007a,Chen2006,Bell2008,Donev2009a,Banchio2003}.  
This includes the Stochastic Immersed Boundary Method~\cite{Atzberger2007a}, 
Fluctuating Immersed Material Dynamics~\cite{Chen2006}, Computational 
Fluctuating Fluid Dynamics~\cite{Bell2008,Donev2009a},
and Accelerated Stokesian Dynamics~\cite{Banchio2003}.  The formalism 
presented here can be regarded in part as a generalization
of these approaches.  It is expected that many of the presented
results can be applied to further justify and validate these
methods and to provide further extensions.  The formalism 
presented here provides a rather general framework for the 
development of computational methods for applications requiring 
a consistent treatment of structure mechanics, hydrodynamic 
coupling, and thermal fluctuations.

%
%
%
%
%
%
%
%
%
%
%

%
%
%
%
%


\section{Summary of the Stochastic Eulerian Lagrangian Method }
\label{sec_summary_SELM}

We summarize here the Stochastic Eulerian Lagrangian Method,
abbreviated as SELM.
We present the general formalism and a number of alternative 
descriptions of the fluid-structure system.  In many situations 
the stochastic differential equations for the full fluid-structure 
dynamics exhibits stiffness.  To cope with this issue and to develop 
efficient numerical methods, simplified descriptions are discussed 
for various physical regimes.  A more detailed discussion and 
derivation of SELM and the reduced equations in each of the 
physical regimes is given in Section~\ref{sec_derivations}.

\begin{figure}[t*]
\centering
\includegraphics[width=5in]{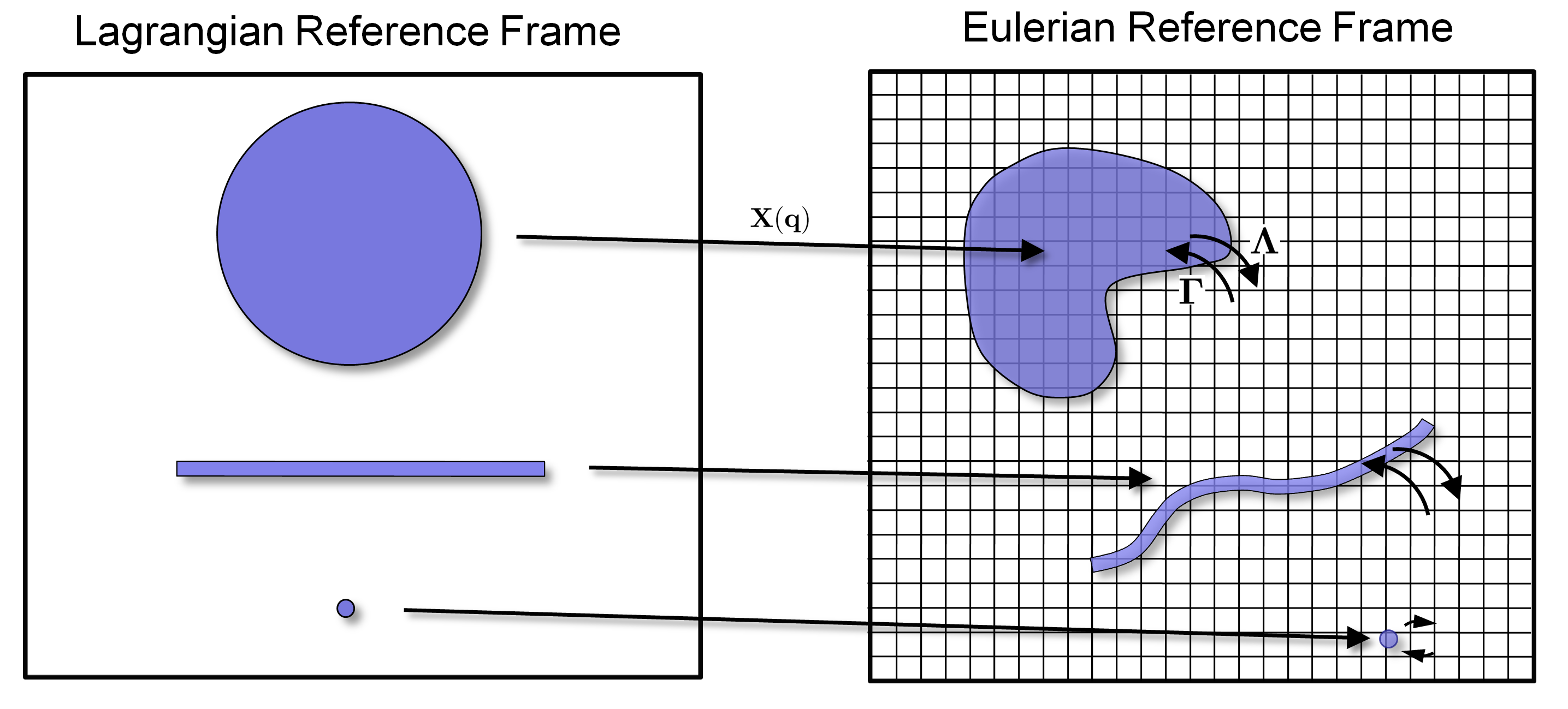}
\caption[Mixed Eulerian and Lagrangian Description]
{The description of the fluid-structure system utilizes both
Eulerian and Lagrangian reference frames.  The structure mechanics
are often most naturally described using a Lagrangian reference 
frame.  The fluid mechanics are often most naturally described
using an Eulerian reference frame.  The mapping $\mathbf{X}(\mathbf{q})$ 
relates the Lagrangian reference frame to the Eulerian reference frame.  
The operator $\OpFSCouple$ prescribes how structures are to be 
coupled to the fluid.  The operator $\OpSFCouple$ prescribes how the 
fluid is to be coupled to the structures.  A variety of fluid-structure
interactions can be represented in this way.  This includes rigid and 
deformable bodies, membrane structures, polymeric structures, or point 
particles.  
}
\label{figure_EL_schematic}
\end{figure}

To study the dynamics of fluid-structure interactions 
in the presence of thermal fluctuations, we utilize a
mechanical description involving Eulerian and Lagrangian 
reference frames.  Such mixed descriptions arise rather naturally, 
since it is often convenient to describe the structure configurations 
in a Lagrangian reference frame while it is convenient to describe
the fluid in an Eulerian reference frame.
In principle more general descriptions using other 
reference frames could also be considered.  Descriptions 
for fluid-structure systems having these features can be described 
rather generally by the following dynamic equations 
\begin{eqnarray}
\label{equ_SELM_0_1}
\rho\frac{d\mathbf{u}}{dt} & = & \OpFldDissp \mathbf{u} + \OpSFCouple[\OpFSDissp(\mathbf{v} - \OpFSCouple\mathbf{u})] + \lambda + \mathbf{f}_\subtxt{thm} \\
\label{equ_SELM_0_2}
m\frac{d\mathbf{v}}{dt}
    & = & -\OpFSDissp\left(\mathbf{v} - \OpFSCouple\mathbf{u}\right)
                            -\nabla_{\mathbf{X}}\Phi[\mathbf{X}]
                            + \zeta
                            + \mathbf{F}_\subtxt{thm} \\
\label{equ_SELM_0_3}
\frac{d\mathbf{X}}{dt}     & = & \mathbf{v}.
\end{eqnarray}
The $\mathbf{u}$ denotes the velocity of the fluid, $\rho$ the uniform fluid density.  The 
$\mathbf{X}$ denotes the configuration of the structure and $\mathbf{v}$ the 
velocity of the structure.  The mass of the structure is denoted by $m$.  
To simplify the presentation we treat here only the case when $\rho$ and $m$ are constant, but 
with some modifications these could also be treated as variable.  
The $\lambda, \zeta$ are Lagrange multipliers for
imposed constraints, such as incompressibility of the fluid or a rigid body constraint of 
a structure.  The operator $\OpFldDissp$ is used to account for dissipation in 
the fluid, such as associated with Newtonian fluid stresses~\cite{Acheson1990}.  
To account for how the fluid and structures are coupled, a few 
general operators are introduced, $\OpFSCouple, \OpFSDissp, \OpSFCouple$.  

The linear operators $\OpFSCouple, \OpSFCouple, \OpFSDissp$ are used to model 
the fluid-structure coupling.  The $\OpFSCouple$ operator describes how a structure depends
on the fluid flow while $-\OpFSDissp$ is a negative definite dissipative operator describing 
the viscous interactions coupling the structure to the fluid.  We assume throughout that this
dissipative operator is symmetric, $\OpFSDissp = \OpFSDissp^{\RegAdjoint}$.  
The linear operator $\OpSFCouple$ is used
to attribute a spatial location for the viscous interactions between the structure and 
fluid.  The linear operators are assumed to have dependence only on the configuration degrees 
of freedom $\OpFSCouple = \OpFSCouple[\mb{X}]$, 
$\OpSFCouple = \OpSFCouple[\mb{X}]$.  We assume further that
$\OpFSDissp$ does not have any dependence on $\mb{X}$.

To account for the mechanics of 
structures, $\Phi[\mathbf{X}]$ denotes the potential energy of the 
configuration $\mathbf{X}$. 
The total energy associated with this fluid-structure system is given by 
\begin{eqnarray}
\label{equ_energy}
E[\mathbf{u}, \mathbf{v},\mathbf{X}] & = & \int_{\Omega}\frac{1}{2}\rho |\mathbf{u}(\mathbf{y})|^2 d\mathbf{y} + \frac{1}{2}m\mathbf{v}^2 + \Phi[\mathbf{X}].
\end{eqnarray}
The first two terms give the kinetic energy of the fluid and structures.  The last term gives
the potential energy of the structures.  

As we shall discuss, it is natural to consider coupling operators $\OpSFCouple$ and
$\OpFSCouple$ which are adjoint in the sense
\begin{eqnarray}
\label{equ_adjoint_cond}
\int_{\mathcal{S}} (\OpFSCouple\mb{u})(\mb{q}) \cdot \mathbf{v}(\mb{q}) d\mb{q} 
= 
\int_{\Omega} \mb{u}(\mb{x}) \cdot (\OpSFCouple\mathbf{v})(\mb{x}) d\mb{x}
\end{eqnarray}
for any $\mathbf{u}$ and $\mathbf{v}$.
The $\mathcal{S}$ and $\Omega$ denote the spaces used to parameterize respectively 
the structures and the fluid.  We denote such an adjoint by
$\OpSFCouple = \OpFSCouple^{\OpAdjoint}$ or $\OpFSCouple = \OpSFCouple^{\OpAdjoint}$.
This adjoint condition can be shown to have the important consequence
that the fluid-structure coupling conserves energy when $\OpFSDissp \rightarrow \infty$
in the inviscid and zero temperature limit.  

In practice, the conditions discussed above can be relaxed somewhat.  For our present purposes 
these conditions help simplify the presentation.  Each of these operators will 
be discussed in more detail below.

To account for thermal fluctuations, a random force density 
$\mathbf{f}_\subtxt{thm}$ is introduced in the fluid equations
and $\mathbf{F}_\subtxt{thm}$ in the structure equations.  These account for 
spontaneous changes in the system momentum which occurs as a result of 
the influence of unresolved microscopic degrees of freedom and 
unresolved events occurring in the fluid and in the fluid-structure 
interactions.

The thermal fluctuations consistent with the form of the total energy and relaxation 
dynamics of the system are taken into account by the introduction of stochastic driving fields in the 
momentum equations of the fluid and structures.  The stochastic driving fields are taken to be Gaussian processes 
with mean zero and with $\delta$-correlation in time~\cite{Reichl1998}.  By the fluctuation-dissipation principle~\cite{Reichl1998} these have 
covariances given by 
\begin{eqnarray}
\label{equ_SELM_thermal1_0}
\langle \mathbf{f}_\subtxt{thm}(s)\mathbf{f}_\subtxt{thm}^{\RegAdjoint}(t) \rangle & = & -\left(2k_B{T}\right)\left(\OpFldDissp - \OpSFCouple\OpFSDissp\OpFSCouple\right)\hspace{0.03cm}\delta(t - s) \\
\label{equ_SELM_thermal2_0}
\langle \mathbf{F}_\subtxt{thm}(s)\mathbf{F}_\subtxt{thm}^{\RegAdjoint}(t) \rangle & = & \left(2k_B{T}\right)\OpFSDissp\hspace{0.03cm}\delta(t - s)\\
\label{equ_SELM_thermal3_0}
\langle \mathbf{f}_\subtxt{thm}(s)\mathbf{F}_\subtxt{thm}^{\RegAdjoint}(t) \rangle & = & -\left(2k_B{T}\right)\OpSFCouple\OpFSDissp\hspace{0.03cm}\delta(t - s).
\end{eqnarray}
We have used that $\OpFSCouple = \OpSFCouple^{\OpAdjoint}$ and $\OpFSDissp = \OpFSDissp^{\RegAdjoint}$.
We remark that the notation $\mb{g}\mb{h}^T$ which is used for the covariance operators 
should be interpreted as the tensor product.  This notation is meant to suggest 
the analogue to the outer-product operation which holds in the discrete 
setting~\cite{Atzberger2010a}.  A more detailed discussion and derivation of the thermal 
fluctuations is given in Section~\ref{sec_derivations}.

It is important to mention that some care must be taken when using the above formalism in practice and when choosing operators.
An important issue concerns the treatment of the material derivative 
of the fluid, $d\mb{u}/dt = \partial \mb{u}/\partial{t} + \mb{u}\cdot\nabla\mb{u}$.  For stochastic systems the field 
$\mb{u}$ is often highly irregular and not defined in a point-wise sense, but rather only in the sense of 
a generalized function (distribution)~\cite{Da1992,Lieb2001}.  This presents issues in how to define the non-linear
term arising in the material derivative, which appears to require point-wise values of $\mb{u}$.
For such irregular velocity fields, this also calls into question the applicability of the theorems 
typically used to derive the differential equations from the conservation laws.  For instance, for such
velocity fields the fluid material body may no longer exhibit smooth deformations over time.

There are a number of ways to deal with this issue.
The first is to consider a regularization of the fluid stresses, which are typically the source of 
irregularity, see equation~\ref{equ_SELM_thermal1_0}.  This can be motivated by 
the fact that the fluid stress tensors typically considered in 
continuum mechanics are expected to become inaccurate at molecular length-scales.  Ideally, from molecular models of 
the fluid the small-length scale (large wave-number) responses of the fluid could be determined 
and provide a justified regularization.  For instance, this could provide an alternative to using responses based on 
Newtonian stresses for all length-scales.  For the SELM formalism, this
would simply correspond to using for $\OpFldDissp$ an alternative to the dissipative operator based 
on Newtonian stresses.  The second more easily implemented approach is simply to work with the 
linearized material derivative, which still retains many of the essential features of the 
fluid dynamics and is useful for many applications~\cite{Atzberger2007a}.  

In this initial presentation of SELM, we shall take the latter approach and treat $d\mb{u}/dt = \partial \mb{u}/\partial{t}$.
This provides a rather general description of fluid-structure systems which incorporate the role of thermal fluctuations.  
From this initial formalism of SELM, we shall derive a number of simplified descriptions for various physical regimes.  
These simplified descriptions for each regime tend to yield less stiff differential equations and have other features
making them useful in the development of efficient stochastic numerical methods for the formalism.

\subsection{Regime I}
We now consider the regime in which the full dynamics of the fluid-structure system are retained but reformulated 
in terms of a field describing the total momentum of the fluid-structure system at 
a given spatial location.  This description is more convenient to work with in practice since 
it results in simplifications in the stochastic driving fields.  For this purpose we define
\begin{eqnarray}
\mathbf{p}(\mathbf{x},t) = \rho \mathbf{u}(\mathbf{x},t) + \OpSFCouple[m\mathbf{v}(t)](\mathbf{x}). 
\end{eqnarray}
The operator $\OpSFCouple$ is used to give the distribution in space of the momentum associated with the structures
for given configuration $\mb{X}(t)$. 
Using this approach, the fluid-structure dynamics are described by
\begin{eqnarray}
\label{equ_SELM_I_1}
\frac{d\mathbf{p}}{dt}  & = & \OpFldDissp \mathbf{u} + \OpSFCouple[-\nabla_{\mathbf{X}}\Phi(\mathbf{X})] 
                         + \left(\nabla_{\mathbf{X}} \OpSFCouple[m\mathbf{v}]\right)\cdot \mathbf{v} 
                         + \lambda + \mathbf{g}_\subtxt{thm} \\
\label{equ_SELM_I_2}
m\frac{d\mathbf{v}}{dt} & = & -\OpFSDissp\left(\mathbf{v} - \OpFSCouple\mathbf{u}\right) 
                            -\nabla_{\mathbf{X}}\Phi(\mathbf{X})
                            + \zeta + \mathbf{F}_\subtxt{thm} \\
\label{equ_SELM_I_3}
\frac{d\mathbf{X}}{dt}  & = & \mathbf{v}                            
\end{eqnarray}
where $\mathbf{u} = \rho^{-1}\left(\mathbf{p} - \OpSFCouple[m\mathbf{v}]\right)$ and 
$\mathbf{g}_\subtxt{thm} = \mathbf{f}_\subtxt{thm} + \OpSFCouple[\mathbf{F}_\subtxt{thm}]$.
The third term in the first equation arises from the dependence of $\OpSFCouple$ on the 
configuration of the structures,
$\OpSFCouple[m\mb{v}] = (\OpSFCouple[X])[m\mb{v}]$.  
The Lagrange multipliers for imposed constraints are denoted by $\lambda, \zeta$.  
For the constraints, we use rather liberally the notation with the 
Lagrange multipliers denoted here not necessarily assumed to be equal to the previous definition.
The stochastic driving fields are again Gaussian with mean zero and $\delta$-correlation in time~\cite{Reichl1998}.  
The stochastic driving fields have the covariance
structure given by 
\begin{eqnarray}
\label{equ_SELM_I_thermal1}
\langle \mathbf{g}_\subtxt{thm}(s)\mathbf{g}_\subtxt{thm}^{\RegAdjoint}(t) \rangle & = & -\left(2k_B{T}\right)\OpFldDissp\hspace{0.06cm}\delta(t - s) \\
\label{equ_SELM_I_thermal2}
\langle \mathbf{F}_\subtxt{thm}(s)\mathbf{F}_\subtxt{thm}^{\RegAdjoint}(t) \rangle & = & \left(2k_B{T}\right)\OpFSDissp\hspace{0.06cm}\delta(t - s) \\
\label{equ_SELM_I_thermal3}
\langle \mathbf{g}_\subtxt{thm}(s)\mathbf{F}_\subtxt{thm}^{\RegAdjoint}(t) \rangle & = & 0.
\end{eqnarray}
This formulation has the convenient feature that the stochastic driving fields become 
independent.  This is a consequence of using the field for the total momentum for which 
the dissipative exchange of momentum between the fluid and structure no longer arises. 
In the equations for the total momentum, the only source of dissipation remaining occurs from 
the stresses of the fluid.  This approach simplifies the effort required to generate 
numerically the stochastic driving fields and will be used throughout.

\subsection{Regime II}  
\label{sec_summary_SELM_regimeII}
We now consider a regime in which the formalism can be simplified significantly.
In many situations, inertial effects often play a relatively minor role in the structure 
dynamics as a consequence of the small mass of the structure relative to the displaced fluid
or as a consequence of viscosity of the solvent fluid~\cite{Hauge1973,Mazur1974}.
In such a regime, the relatively rapid dynamics associated with the momentum of the structures often 
presents a source of stiffness in numerical calculations.  To cope with these issues we consider 
a reduction of 
the stochastic dynamics of the system in which the structure momentum is 
eliminated from the description.  In particular, we consider the regime in
which $m \ll \rho \ell^3$.  The $\ell$ denotes a length-scale characteristic of the 
size of the immersed structure and associated flow field of the fluid.
In the limit $m \rightarrow 0$, the fluid-structure dynamics are governed by
\begin{eqnarray}
\label{equ_SELM_II_1}
\frac{d\mathbf{p}}{dt}  & = & \rho^{-1}\OpFldDissp \mathbf{p} + \OpSFCouple[-\nabla_{\mathbf{X}}\Phi(\mathbf{X})] 
                         + \left(\nabla_{\mathbf{X}}\cdot\OpSFCouple\right) k_B{T} + \lambda + \mathbf{g}_\subtxt{thm} \\
\label{equ_SELM_II_2}
\frac{d\mathbf{X}}{dt}  & = & \rho^{-1}\OpFSCouple\mathbf{p} + \OpFSDissp^{-1}[-\nabla_{\mathbf{X}}\Phi(\mathbf{X})] + \zeta + \mathbf{G}_\subtxt{thm}.
\end{eqnarray}
In the notation $\nabla_{\mathbf{X}}\cdot\OpSFCouple = \mbox{Tr}[\nabla_{\mathbf{X}}\OpSFCouple]$. 
This term arises from the thermal fluctuations associated with the 
momentum of the structures, which have been eliminated from the description.  
This term plays an important role in the system when the 
phase-space dynamics of $(\mathbf{p},\mathbf{X})$ has an associated
vector field which is compressible.  When considering the Liouville 
equation on phase-space, this term accounts for local changes 
of the phase-space volume which occurs as the configuration of 
the structure changes under the dynamics of the reduced description~\cite{Tuckerman1999}.  
For a more detailed discussion see Section~\ref{sec_equil_stat_mech_regime_II}.
The stochastic driving fields $\mathbf{g}_\subtxt{thm},\mathbf{G}_\subtxt{thm}$
are again Gaussian with mean zero and with $\delta$-correlation in time~\cite{Reichl1998}.  The 
stochastic driving fields have the covariance structure given by 
\begin{eqnarray}
\langle \mathbf{g}_\subtxt{thm}(s)\mathbf{g}_\subtxt{thm}^{\RegAdjoint}(t) \rangle & = & -\left(2k_B{T}\right)\OpFldDissp\hspace{0.06cm}\delta(t - s) \\
\langle \mathbf{G}_\subtxt{thm}(s)\mathbf{G}_\subtxt{thm}^{\RegAdjoint}(t) \rangle & = & \left(2k_B{T}\right)\OpFSDissp^{-1}\hspace{0.06cm}\delta(t - s) \\
\langle \mathbf{g}_\subtxt{thm}(s)\mathbf{G}_\subtxt{thm}^{\RegAdjoint}(t) \rangle & = & 0. 
\end{eqnarray}
A more detailed discussion and derivation of the equations in this regime is given in 
Section~\ref{sec_derivation_regime_II}.

\subsection{Regime III}
\label{sec_summary_SELM_regimeIII}
The description of the fluid-structure system can be further simplified by considering
the viscous coupling between the fluid and structures in the limit 
$\OpFSDissp \rightarrow \infty$.  In this case, the fluid-structure dynamics are 
given by
\begin{eqnarray}
\label{equ_SELM_III_1}
\frac{d\mathbf{p}}{dt}  & = & \rho^{-1}\OpFldDissp \mathbf{p} + \OpSFCouple[-\nabla_{\mathbf{X}}\Phi(\mathbf{X})] 
                        + \left(\nabla_{\mathbf{X}}\cdot\OpSFCouple\right) k_B{T} + \lambda + \mathbf{g}_\subtxt{thm} \\
\label{equ_SELM_III_2}
\frac{d\mathbf{X}}{dt}  & = & \rho^{-1}\OpFSCouple\mathbf{p} \\
\label{equ_SELM_III_3}
\langle \mathbf{g}_\subtxt{thm}(s)\mathbf{g}_\subtxt{thm}^{\RegAdjoint}(t) \rangle & = & -\left(2k_B{T}\right)\OpFldDissp\hspace{0.06cm}\delta(t - s).
\end{eqnarray}
In the notation $\nabla_{\mathbf{X}}\cdot\OpSFCouple = \mbox{Tr}[\nabla_{\mathbf{X}}\OpSFCouple]$. 
A more detailed discussion and derivation of the equations in this regime in given in 
Section~\ref{sec_derivation_regime_III}.

\subsection{Regime IV}
\label{sec_summary_SELM_regimeIV}
The description of the fluid-structure system can be further simplified by considering
for the fluid the viscous limit in which $\mu \rightarrow \infty$.  In this regime
only the structure dynamics remain and can be shown to be given by 
\begin{eqnarray}
\label{equ_SELM_IV_1}
\frac{d\mathbf{X}}{dt}  & = & H_\subtxt{SELM}[-\nabla_{\mathbf{X}}\Phi(\mathbf{X})] 
+ (\nabla_{\mb{X}}\cdot H_\subtxt{SELM})k_B{T}
+ \mathbf{h}_\subtxt{thm} \\
\label{equ_SELM_IV_2}
H_\subtxt{SELM}         & = & \OpFSCouple(-\wp\OpFldDissp)^{-1}\OpSFCouple \\
\label{equ_SELM_IV_3}
\langle \mathbf{h}_\subtxt{thm}(s)\mathbf{h}_\subtxt{thm}^{\RegAdjoint}(t) \rangle & = & \left(2k_B{T}\right)H_\subtxt{SELM}\hspace{0.06cm}\delta(t - s).
\end{eqnarray}
The $\wp$ denotes a projection operator imposing constraints, 
such as incompressibility.  The adjoint property   
$\OpSFCouple = \OpFSCouple^{\OpAdjoint}$
and symmetry of $\wp\OpFldDissp$ yields an 
operator $H_\subtxt{SELM}$ which is symmetric.  
A more detailed discussion and derivation of the equations in this regime is given in 
Section~\ref{sec_derivation_regime_IV}.

\subsection{Summary}
This gives an overview of the SELM formalism and the associated 
stochastic differential equations.  We remark that each of these regimes were 
motivated by a rather specific limit.  Non-dimensional analysis 
of the equations can also be carried out and other limits considered to motivate working with such 
reduced equations.  We discuss in more detail the derivation of the reduced equations in each regime 
in Section~\ref{sec_derivations}.  We discuss how specific stochastic numerical methods can be 
developed for the SELM formalism in Section~\ref{sec_comp_method}.  We discuss applications and
how the SELM formalism can be used in practice in Section~\ref{sec_applications}.   

\section{Derivations for the Stochastic Eulerian Lagrangian Method}
\label{sec_derivations}

We now discuss formal derivations to motivate the 
stochastic differential equations used in each of the physical 
regimes.  For this purpose, we do not present the most general derivation of 
the equations.  For brevity, we make simplifying assumptions when convenient.

In the initial formulation of SELM, the fluid-structure system is described by
\begin{eqnarray}
\label{SELM_equ_deriv1_0}
\rho\frac{d\mathbf{u}}{dt} & = & \OpFldDissp \mathbf{u} + \OpSFCouple[\OpFSDissp(\mathbf{v} - \OpFSCouple\mathbf{u})] + \lambda
                            + \mathbf{f}_\subtxt{thm} \\
\label{SELM_equ_deriv2_0}
m\frac{d\mathbf{v}}{dt}
    & = & -\OpFSDissp\left(\mathbf{v} - \OpFSCouple\mathbf{u}\right)
                            -\nabla_{\mathbf{X}}\Phi(\mathbf{X})
                            + \zeta
                            + \mathbf{F}_\subtxt{thm} \\
\label{SELM_equ_deriv3_0}
\frac{d\mathbf{X}}{dt}     & = & \mathbf{v}.                            
\end{eqnarray}
The notation and operators appearing in these equations has been discussed in detail in 
Section~\ref{sec_summary_SELM}.
For these equations, we focus primarily on the motivation for the stochastic driving fields 
used for the fluid-structure system.  

For the thermal fluctuations of the system, we assume Gaussian random fields with
mean zero and $\delta$-correlated in time.  For such stochastic fields, the 
central challenge is to determine an appropriate covariance structure.  
For this purpose, we use the fluctuation-dissipation principle of 
statistical mechanics~\cite{Reichl1998, LandauStatMech1980}.
For linear stochastic differential equations of the form
\begin{eqnarray}
d\mathbf{Z}_t = L\mathbf{Z}_t dt + Q d\mathbf{B}_t
\end{eqnarray}
the fluctuation-dissipation principle can be expressed as
\begin{eqnarray}
\label{equ_deriv_FLD}
G = QQ^{\RegAdjoint} = -(LC) - (LC)^{\RegAdjoint}. 
\end{eqnarray}
This relates the equilibrium covariance structure $C$ of the system to the covariance structure
$G$ of the stochastic driving field.  The operator $L$ accounts for the dissipative
dynamics of the system.  
For the equations~\ref{SELM_equ_deriv1_0} -- ~\ref{SELM_equ_deriv3_0}, the dissipative 
operators only appear in the momentum equations.  This can be shown to have the 
consequence that there is no
thermal forcing in the equation for $\mathbf{X}(t)$, this will also be confirmed in 
Section~\ref{sec_equil_stat_mech_regime_I}.  To simplify the presentation, we do not 
represent explicitly the stochastic dynamics of the structure configuration 
$\mathbf{X}$.

For the fluid-structure system it is convenient to work with the stochastic driving fields by defining
\begin{eqnarray}
\mathbf{q} & = & [\rho^{-1}\mathbf{f}_\subtxt{thm},m^{-1}\mathbf{F}_\subtxt{thm}]^{\RegAdjoint}.
\end{eqnarray}
The field $\mathbf{q}$ formally is given by $\mathbf{q} = Q d\mathbf{B}_t/dt$ and determined by the 
covariance structure $G =QQ^{\RegAdjoint}$.  This covariance 
structure is determined by the fluctuation-dissipation principle expressed in equation~\ref{equ_deriv_FLD} with
\begin{eqnarray}
L                                      & = & \left\lbrack
\begin{array}{ll}
\rho^{-1}\left(\OpFldDissp - \OpSFCouple\OpFSDissp\OpFSCouple \right) & \rho^{-1} \OpSFCouple \OpFSDissp \\
m^{-1} \OpFSDissp\OpFSCouple & -m^{-1} \OpFSDissp \\
\end{array}
\right\rbrack \\
C                                      & = & \left\lbrack
\begin{array}{ll}
\rho^{-1} k_B{T}\mathcal{I} & 0 \\
0 & m^{-1} k_B{T}\mathcal{I} \\
\end{array}
\right\rbrack. 
\end{eqnarray}
The $\mathcal{I}$ denotes the identity operator.
The covariance $C$ was obtained by considering the fluctuations at equilibrium.  
The covariance $C$ is easily found since the Gibbs-Boltzmann distribution 
is a Gaussian with formal density 
$\Psi(\mathbf{u},\mathbf{v}) = \frac{1}{Z_0} \exp\left[-E/k_B{T}\right]$.
The $Z_0$ is the normalization constant for $\Psi$.  The energy is given 
by equation~\ref{equ_energy}.  For this purpose, we need only 
consider the energy $E$ in the case when $\Phi = 0$.    This gives the covariance structure 
\begin{eqnarray}
G                                      & = & \left(2k_BT\right)\left\lbrack
\begin{array}{ll}
-\rho^{-2}\left(\OpFldDissp - \OpSFCouple\OpFSDissp\OpFSCouple \right) & -m^{-1}\rho^{-1} \OpSFCouple \OpFSDissp \\
-m^{-1}\rho^{-1} \OpFSDissp\OpFSCouple & m^{-2} \OpFSDissp \\
\end{array}
\right\rbrack .
\end{eqnarray}
To obtain this result we use that $\OpFSCouple = \OpSFCouple^{\OpAdjoint}$ and $\OpFSDissp = \OpFSDissp^{\OpAdjoint}$.
From the definition of $\mathbf{q}$, it is found the covariance of the stochastic driving fields of SELM are
given by equations~\ref{equ_SELM_thermal1_0}--~\ref{equ_SELM_thermal3_0}.
This provides a description of the thermal fluctuations in the fluid-structure system.

\subsection{Regime I}
\label{sec_derivation_regime_I}
It is convenient to reformulate the description of the fluid-structure system
in terms of a field for the total momentum of the system associated with spatial location $\mathbf{x}$.  
For this purpose we define
\begin{eqnarray}
\mathbf{p}(\mathbf{x},t) = \rho \mathbf{u}(\mathbf{x},t) + \OpSFCouple[m\mathbf{v}(t)](\mathbf{x}). 
\end{eqnarray}
The operator $\OpSFCouple$ is used to give the distribution in space of the momentum associated with the structures. 
Using this approach, the fluid-structure dynamics are described by
\begin{eqnarray}
\frac{d\mathbf{p}}{dt}  & = & \OpFldDissp \mb{u} + \OpSFCouple[-\nabla_{\mathbf{X}}\Phi(\mathbf{X})] 
                         + (\nabla_{\mathbf{X}} \OpSFCouple[m\mathbf{v}])\cdot \mathbf{v} + \lambda + \mathbf{g}_\subtxt{thm} \\
m\frac{d\mathbf{v}}{dt} & = & -\OpFSDissp\left(\mathbf{v} - \OpFSCouple\mathbf{u}\right) 
                            -\nabla_{\mathbf{X}}\Phi(\mathbf{X})
                            + \zeta
                            + \mathbf{F}_\subtxt{thm} \\
\frac{d\mathbf{X}}{dt}  & = & \mathbf{v}                
\end{eqnarray}
where $\mathbf{u} = \rho^{-1}\left(\mathbf{p} - \OpSFCouple[m\mathbf{v}]\right)$ 
and $\mathbf{g}_\subtxt{thm} = \mathbf{f}_\subtxt{thm} + \OpSFCouple[\mathbf{F}_\subtxt{thm}]$.
The third term in the first equation arises from the dependence of $\OpSFCouple$ on the 
configuration of the structures, $\OpSFCouple[m\mathbf{v}(t)] = (\OpSFCouple[X])[m\mathbf{v}(t)]$. 

The thermal fluctuations are taken into account by two stochastic fields 
$\mathbf{g}_\subtxt{thm}$ and $\mathbf{F}_\subtxt{thm}$.  
The covariance of $\mathbf{g}_\subtxt{thm}$ is obtained from
\begin{eqnarray}
\langle \mathbf{g}_\subtxt{thm}\mathbf{g}_\subtxt{thm}^{\RegAdjoint} \rangle & = & 
\langle \mathbf{f}_\subtxt{thm}\mathbf{f}_\subtxt{thm}^{\RegAdjoint} \rangle
+
\langle \mathbf{f}_\subtxt{thm}\mathbf{F}_\subtxt{thm}^{\RegAdjoint} \OpSFCouple^{\RegAdjoint} \rangle
+
\langle \OpSFCouple\mathbf{F}_\subtxt{thm}\mathbf{f}_\subtxt{thm}^{\RegAdjoint} \rangle
+
\langle \OpSFCouple\mathbf{F}_\subtxt{thm}\mathbf{F}_\subtxt{thm}^{\RegAdjoint}\OpSFCouple^{\RegAdjoint} \rangle \\
\nonumber
& = & 
(2k_B{T})\left(-\OpFldDissp + \OpSFCouple \OpFSDissp \OpFSCouple 
- \OpSFCouple \OpFSDissp \OpSFCouple^{\RegAdjoint} - \OpSFCouple \OpFSDissp \OpSFCouple^{\RegAdjoint}
+ \OpSFCouple \OpFSDissp \OpSFCouple^{\RegAdjoint}
\right) \\
\nonumber
& = & 
-\left(2k_B{T}\right)\OpFldDissp.
\end{eqnarray}
This makes use of the adjoint property of the coupling operators
$\OpSFCouple^{\OpAdjoint} = \OpFSCouple$. 

One particularly convenient feature of this reformulation is that the stochastic driving field
$\mathbf{F}_\subtxt{thm}$ and $\mathbf{g}_\subtxt{thm}$ become independent.  This can 
be seen as follows
\begin{eqnarray}
\langle \mathbf{g}_\subtxt{thm}\mathbf{F}_\subtxt{thm}^{\RegAdjoint} \rangle & = & 
\langle \mathbf{f}_\subtxt{thm} \mathbf{F}_\subtxt{thm}^{\RegAdjoint} \rangle
+
\langle \OpSFCouple \mathbf{F}_\subtxt{thm} \mathbf{F}_\subtxt{thm}^{\RegAdjoint} \rangle \\
\nonumber
& = & (2k_B{T})(-\OpSFCouple\OpFSDissp  +  \OpSFCouple\OpFSDissp) = 0.
\end{eqnarray}
This decoupling of the stochastic driving fields greatly reduces the computational 
effort to generate the fields with the required covariance structure.  This shows the 
covariance structure of the stochastic driving fields of SELM are given by 
equations~\ref{equ_SELM_I_thermal1}--~\ref{equ_SELM_I_thermal3}.

\subsection{Regime II}
\label{sec_derivation_regime_II}
In many situations, inertial effects often play a relatively minor role in the structure 
dynamics as a consequence of the small mass of the structure relative to the displaced fluid
or as a consequence of the large viscosity of the solvent fluid~\cite{Hauge1973,Mazur1974}.  
We consider the regime in which $m \ll \rho \ell^3$, as discussed in 
Section~\ref{sec_summary_SELM_regimeII}.  We shall derive formally reduced 
stochastic equations in the limit $m \rightarrow 0$. 

For this purpose, we focus primarily on the dynamics of the velocity of the structures
$\mathbf{v}(t)$.  This can be expressed using the notation of Ito Stochastic Differential 
Equations~\cite{Oksendal2000} as
\begin{eqnarray}
\label{equ_deriv_V_t}
d\mathbf{V}_t & = & -m^{-1}\OpFSDissp\left(\mathbf{V}_t - \OpFSCouple\mathbf{u}
                            +\OpFSDissp^{-1}\nabla_{\mathbf{X}}\Phi(\mathbf{X})
                                     \right)dt
                            + m^{-1}(2 k_B{T}\OpFSDissp)^{1/2} d\mathbf{B}_t.
\end{eqnarray}
The $\mathbf{B}_t$ denotes throughout the standard Brownian motion on $\mathbb{R}^N$~\cite{Oksendal2000}.
To simplify the presentation, we consider the case when $\zeta = 0$.  We expect similar results
to hold more generally.
Treating the other degrees of freedom as fixed, we can solve equation~\ref{equ_deriv_V_t} using 
Ito's Lemma~\cite{Oksendal2000}.  The stationary behavior of this stochastic process can be expressed as
\begin{eqnarray}
\mathbf{V}_t       & = & \boldsymbol{\mu}_0 + \int_{-\infty}^t e^{-(t - s)m^{-1}\OpFSDissp} m^{-1} (2 k_B{T}\OpFSDissp)^{1/2} d\mathbf{B}_s \\
\boldsymbol{\mu}_0 & = & \OpFSCouple\mathbf{u} -\OpFSDissp^{-1}\nabla_{\mathbf{X}}\Phi(\mathbf{X}).
\end{eqnarray}
As a result of the integrand being deterministic in the Ito Integral, the $\mathbf{V}_t$ is a Gaussian process.  This
has the consequence that the statistics of the process $\mathbf{V}_t$ are completely determined 
by its mean and covariance functions.  The mean of the process is given at each time by 
\begin{eqnarray}
\boldsymbol{\mu}(t) =  \langle \mathbf{V}_t \rangle = \boldsymbol{\mu}_0.
\end{eqnarray}
The covariance function can be computed using the Ito Isometry~\cite{Oksendal2000} to obtain 
\begin{eqnarray}
\phi(|\tau|) & = & \langle (\mathbf{V}_{t + \tau} - \boldsymbol{\mu}_0) (\mathbf{V}_t - \boldsymbol{\mu}_0)^{\RegAdjoint} \rangle 
 =  k_B{T} m^{-1} e^{-|\tau|m^{-1}\OpFSDissp } \mathcal{I}.
\end{eqnarray}
In the limit $m \rightarrow 0$ this can be expressed as
\begin{eqnarray}
\phi(|t - s|) = {2k_B{T}}\OpFSDissp^{-1} \left[\frac{1}{2} m^{-1}\OpFSDissp  e^{-|t - s|m^{-1}\OpFSDissp }\right]
            \rightarrow {2k_B{T}}\OpFSDissp^{-1} \delta(t - s).
\end{eqnarray}
We have used formally $\frac{1}{2}\lambda e^{-\lambda|\tau|} \rightarrow \delta(\tau)$ as $\lambda \rightarrow \infty$.
This suggests the following approximation for the velocity of the structures in equations~\ref{equ_SELM_I_2} and~\ref{equ_SELM_I_3}.
\begin{eqnarray}
\mathbf{v}(t) & \rightarrow & \OpFSCouple\mathbf{u} -\OpFSDissp^{-1}\nabla_{\mathbf{X}}\Phi(\mathbf{X}) 
+ \left({2k_B{T}}{\OpFSDissp^{-1}}\right)^{1/2} \frac{d\mathbf{B}_t}{dt}. 
\end{eqnarray}

To approximate the term $(\nabla_{\mathbf{X}} \OpSFCouple[m\mathbf{v}])\cdot \mathbf{v}$ in the limit $m \rightarrow 0$ appearing in equation~\ref{equ_SELM_I_1}, 
a different approach is required.  For this purpose, we consider the process $R_t = f(\mathbf{V}_t) = m\mb{V}_t \mb{V}_t^{\RegAdjoint}$.
By Ito's Lemma this satisfies the stochastic differential equation
\begin{eqnarray}
dR_t & = & \nabla f(\mb{V}_t) d\mb{V}_t + \frac{1}{2} d\mb{V}_t^{\RegAdjoint}\nabla^2f(\mb{V}_t) d\mb{V}_t
\end{eqnarray}
with the formal substitutions $dt\hspace{0.05cm}d\mb{B}_t = 0 = dt\hspace{0.05cm}dt$, $d\mb{B}_t\hspace{0.05cm}d\mb{B}_t^{\RegAdjoint} = \mathcal{I} dt$. 
To simplify the discussion we consider the case when $\OpFSDissp = \gamma \mathcal{I}$. We expect similar results can be obtained more generally.
In this case
\begin{eqnarray}
dR_t & = & -2m^{-1}\gamma (R_t - k_B{T} I - \frac{1}{2}(m\mathbf{V}_t\boldsymbol{\mu}_0^{\RegAdjoint} + 
\boldsymbol{\mu}_0(m\mathbf{V}_t)^{\RegAdjoint})) dt \\
\nonumber
& + & m^{-1}(2k_B{T}\gamma)^{1/2}\left(m\mathbf{V}_t d\mathbf{B}_t^{\RegAdjoint} +  d\mathbf{B}_t (m\mathbf{V}_t)^{\RegAdjoint}\right).
\end{eqnarray}
From the form of the energy given by equation~\ref{equ_energy}, the structure momentum $m\mathbf{V}_t$ has 
equilibrium distribution $\Psi(m\mathbf{V}) = \exp\left[{-{(m\mathbf{V})^2}/{2 m k_B{T}}}\right]$.  This gives a 
Gaussian with mean and variance 
\begin{eqnarray}
\langle m\mathbf{V}_t \rangle               & = & 0 \\
\langle (m\mathbf{V}_t)(m\mathbf{V}_t)^{\RegAdjoint} \rangle & = & mk_B{T}\mathcal{I}. 
\end{eqnarray}
By reasoning similar to the arguments used to approximate $\mathbf{V}_t$,
this suggests the terms involving $m\mathbf{V}_t$ do not make a contribution 
in the $m \rightarrow 0$ limit.  This suggests to leading order we have
\begin{eqnarray}
dR_t & = & -2m^{-1}\gamma (R_t - k_B{T} \mathcal{I} )dt.
\end{eqnarray}
This suggests the approximation in the limit $m \rightarrow 0$ 
\begin{eqnarray}
R_t = m\mb{V}_t\mb{V}_t^{\RegAdjoint} \rightarrow k_B{T} \mathcal{I}.
\end{eqnarray}
By substituting this result for $\mb{v}$ in equation~\ref{equ_SELM_I_1}, we have 
\begin{eqnarray}
(\nabla_{\mathbf{X}} \OpSFCouple[m\mathbf{v}])\cdot \mathbf{v} \rightarrow 
\mbox{Tr}[\nabla_{\mathbf{X}} \OpSFCouple](k_B{T}) = (\nabla_{\mathbf{X}}\cdot \OpSFCouple)(k_B{T}).
\end{eqnarray}

This establishes formally the reduced SELM description when the mass of 
the structures becomes negligible.  It should be mentioned these approximations can be established more rigorously 
using a perturbation analysis of the Kolomogorov Equations associated with the stochastic 
processes~\cite{Kramer2003,Gardiner1985}.  It should also be mentioned that other limits can be considered in which additional 
drift and stochastic terms arise in the reduced momentum equations.  This will be the focus of another paper.

\subsection{Regime III}
\label{sec_derivation_regime_III}
The SELM stochastic equations can be further reduced if the effective viscous interactions between the 
structure and fluid are assumed to become very strong.  This corresponds to approximating
the reduced stochastic equations of Regime II in the formal limit 
$\OpFSDissp \rightarrow \infty$.  By this notation, we mean that all eigenvalues 
of the symmetric operator $\OpFSDissp$ uniformly tend to infinity.  In this formal 
limit the terms involving $\OpFSDissp^{-1}$ are expected to no longer make a 
contribution to the dynamics.  This motivates the reduced stochastic 
equations~\ref{equ_SELM_III_1}-~\ref{equ_SELM_III_3}.

\subsection{Regime IV}
\label{sec_derivation_regime_IV}
The description of the fluid-structure system can be further simplified by considering
for the fluid the viscous limit in which $\mu \rightarrow \infty$.  In this regime
the fluid adopts a quasi-steady-state behavior with respect to the configuration of 
the structures and the forces acting on the fluid.  In this limit only 
the structure dynamics remain.  By approximating $\mathbf{u}(t)$ using arguments
similar to those used in Regime II for approximating $\mathbf{V}_t$, we can 
derive the reduced stochastic 
equations~\ref{equ_SELM_IV_1}--~\ref{equ_SELM_IV_3}.

\section{Computational Methodology} 
\label{sec_comp_method}
We now discuss briefly numerical methods for the SELM formalism.
For concreteness we consider the specific case in which the
fluid is Newtonian and incompressible.  For now, the 
other operators of the SELM formalism will be treated 
rather generally.  This case 
corresponds to the dissipative operator for the fluid
\begin{eqnarray}
\OpFldDissp \mathbf{u} = \mu \Delta \mathbf{u}.
\end{eqnarray}
The $\Delta$ denotes the Laplacian 
$\Delta \mathbf{u} = \partial_{xx}\mathbf{u} + \partial_{yy}\mathbf{u} + \partial_{zz}\mathbf{u}$.
The incompressibility of the fluid corresponds to the constraint
\begin{eqnarray}
\nabla \cdot \mathbf{u} = 0.
\end{eqnarray}
This is imposed by the Lagrange multiplier $\lambda$.  By the Hodge Decomposition,
$\lambda$ is given by the gradient of a function $p$ with $\lambda = -\nabla{p}$.  The $p$ can
be interpreted as the local pressure of the fluid.  

A variety of methods could be used in practice to discretize the SELM formalism, 
such as Finite Difference Methods, Spectral Methods, and Finite Element Methods~\cite{Gottlieb1993,Strikwerda2004,Strang2008}.  
We present here discretizations based on Finite Difference Methods.

\subsection{Numerical Semi-Discretizations for Incompressible Newtonian Fluid}
\label{sec_num_method_newtonian}
The Laplacian will be approximated by central differences on a uniform periodic 
lattice by 
\begin{eqnarray}
\left[L\mathbf{u}\right]_{\mb{m}} = \sum_{j = 1}^3 \frac{\mb{u}_{\mb{m} + \mb{e}_j} - 2\mb{u}_{\mb{m}} + \mb{u}_{\mb{m} - \mb{e}_j}}{\Delta{x}^2}.
\end{eqnarray}
The $\mathbf{m} = (m_1, m_2, m_3)$ denotes the index of the lattice site.
The $\mb{e}_j$ denotes the standard basis vector in three dimensions.  
The incompressibility of the fluid will be approximated by imposing the constraint 
\begin{eqnarray}
\left[D\cdot\mathbf{u}\right]_{\mb{m}} = \sum_{j = 1}^3 \frac{\mb{u}_{\mb{m} + \mb{e}_j}^j - \mb{u}_{\mb{m} - \mb{e}_j}^j}{2\Delta{x}}.
\end{eqnarray}
The superscripts denote the vector component.
In practice, this will be imposed by computing the projection of a vector $\mathbf{u}^*$ to the sub-space 
$\{\mathbf{u} \in \mathbb{R}^{3N} \hspace{0.025cm} | \hspace{0.125cm} D\cdot\mathbf{u} = 0 \}$, where 
$N$ is the total number of lattice sites.  We denote this projection operation by 
\begin{eqnarray}
\mathbf{u} = \wp \mathbf{u}^*.
\end{eqnarray}
The semi-discretized equations for SELM to be used in practice are 
\begin{eqnarray}
\label{equ_SELM_semidiscr1}
\frac{d\mathbf{p}}{dt} & = & 
\OpFldDisspDiscr\mb{u} 
+
\OpSFCoupleDiscr[-\nabla_{\mb{X}} \Phi] 
+ 
(\nabla_{\mb{X}}\OpSFCoupleDiscr[m\mb{v}]) \cdot \mb{v}
+ 
\lambda
+ 
\mb{g}_\subtxt{thm}\\
\label{equ_SELM_semidiscr2}
\frac{d\mathbf{v}}{dt} & = & 
-\OpFSDisspDiscr [\mb{v} - \OpFSCoupleDiscr \mb{u}] + \mb{F}_\subtxt{thm} \\
\label{equ_SELM_semidiscr3}
\frac{d\mathbf{X}}{dt} & = & \mb{v}.
\end{eqnarray}
The component $\mb{u}_{\mb{m}} = \rho^{-1}(\mathbf{p}_{\mb{m}} - \OpSFCoupleDiscr[m\mb{v}]_{\mb{m}})$.
Each of the operators now appearing are understood to be discretized.  
We discuss specific discretizations for $\OpFSCoupleDiscr$ and $\OpSFCoupleDiscr$ 
in Section~\ref{sec_applications}.  
To obtain the Lagrange multiplier $\lambda$ which imposes incompressibility we use the projection 
operator and
\begin{eqnarray}
\lambda = 
-(\mathcal{I} - \wp)
\left(
\OpFldDisspDiscr \mb{u}
+ 
\OpFSDisspDiscr [\mb{v} - \OpFSCoupleDiscr \mb{u}]
+ \mb{f}_\subtxt{thm}
\right)
\end{eqnarray}
In this expression, we let $\mb{f}_\subtxt{thm} = \mb{g}_\subtxt{thm} - \OpSFCoupleDiscr[\mb{F}_\subtxt{thm}]$
for the particular realized values of the fields $\mb{g}_\subtxt{thm}$ and $\mb{F}_\subtxt{thm}$.

We remark that in fact the semi-discretized equations of the SELM formalism in this regime can
also be given in terms of $\mb{u}$ directly, which may provide a simpler approach in practice.  
The identity $\mb{f}_\subtxt{thm} = \mb{g}_\subtxt{thm} - \OpSFCoupleDiscr[\mb{F}_\subtxt{thm}]$
could be used to efficiently generate the required stochastic driving fields in the equations
for $\mb{u}$.  We present the reformulation here, since it more directly suggests the 
semi-discretized equations to be used for the reduced stochastic equations.

For this semi-discretization, we consider a total energy for the system given by
\begin{eqnarray}
\label{equ_energy_discr}
E[\mb{u},\mb{v},\mb{X}] = \frac{\rho}{2} \sum_{\mb{m}} |\mb{u}(\mb{x}_{\mb{m}})|^2 \Delta\mb{x}_{\mb{m}}^3 
+ \frac{m}{2} |\mb{v}|^2 + \Eng[\mb{X}].
\end{eqnarray}
This is useful in formulating an adjoint condition~\ref{equ_adjoint_cond} for the 
semi-discretized system.  This can be derived by considering the requirements on 
the coupling operators $\OpFSCoupleDiscr$ and $\OpSFCoupleDiscr$ which ensure the 
energy is conserved when $\OpFSDisspDiscr \rightarrow \infty$ in the inviscid 
and zero temperature limit.

To obtain appropriate behaviors for the thermal fluctuations, it is important
to develop stochastic driving fields which are tailored to the specific semi-discretizations 
used in the numerical methods.  Once the stochastic driving fields are determined, 
which is the subject of the next section, the equations can be integrated in 
time using traditional methods for SDEs, such as the Euler-Maruyama Method or a
Stochastic Runge-Kutta Method~\cite{Platen1992}.  More 
sophisticated integrators in time can also be developed to cope with sources of 
stiffness, but are beyond the scope of this paper~\cite{Atzberger2007a}.  
For each of the reduced equations, similar semi-discretizations can be 
developed as the one presented above.

\subsection{Stochastic Driving Fields for Semi-Discretizations}
\label{sec_gen_stoch_fields}
To obtain behaviors consistent with statistical mechanics, 
it is important stochastic driving 
fields be used which are tailored to the specific numerical discretization
employed~\cite{Atzberger2007a, Donev2009a, Atzberger2010a}.  
To ensure consistency with statistical mechanics, we will again
use the fluctuation-dissipation principle but now apply it to the 
semi-discretized equations.  For each regime, we then discuss the important 
issues arising in practice concerning the efficient generation 
of these stochastic driving fields. 

\subsection{Regime I}
\label{sec_gen_stoch_fields_regime_I}
To obtain the covariance structure for this regime, we apply 
the fluctuation-dissipation principle as expressed in equation~\ref{equ_deriv_FLD}
to the semi-discretized equations~\ref{equ_SELM_semidiscr1}--~\ref{equ_SELM_semidiscr3}.
This gives the covariance 
\begin{eqnarray}
\label{equ_G_regime_I}
G = -2LC = 
(2k_B{T})
\left[
\begin{array}{lll}
-\rho^{-2} \Delta{x}^{-3} \OpFldDisspDiscr &                        0 & 0 \\
0                                          & m^{-2} \OpFSDisspDiscr   & 0 \\
0                                          & 0                        & 0 
\end{array}
\right].
\end{eqnarray}
The factor of $\Delta{x}^{-3}$ arises from the form of the energy for the 
discretized system which gives covariance for the equilibrium fluctuations
of the total momentum $\rho^{-1}\Delta{x}^{-3}k_B{T}$, see equation~\ref{equ_energy_discr}.  
In practice, achieving the covariance associated with the dissipative
operator of the fluid $\OpFldDisspDiscr$ is typically the most challenging to
generate efficiently.  This arises from the large number $N$ of 
lattice sites in the discretization.  

One approach is to determine a factor $Q$ such 
that the block $G_{\mb{p},\mb{p}} = QQ^{\RegAdjoint}$, subscripts 
indicate block entry of the matrix.  The required random 
field with covariance $G_{\mb{p},\mb{p}}$ is then given by 
$\mathbf{g} = Q\boldsymbol{\xi}$,
where $\boldsymbol{\xi}$ is the uncorrelated Gaussian field with 
the covariance structure $\mathcal{I}$.  For the discretization 
used on the uniform periodic mesh, the matrices $L$ and $C$ 
are cyclic~\cite{Strang1988}.  This has the important consequence that 
they are both diagonalizable in the discrete Fourier basis of
the lattice.  As a result, the field $\mathbf{f}_\subtxt{thm}$ 
can be generated using the Fast Fourier Transform (FFT) 
with at most $O(N\log(N))$ computational steps.  In fact, in this 
special case of the discretization, ``random fluxes'' at the cell 
faces can be 
used to generate the field in $O(N)$ computational 
steps~\cite{Atzberger2010a}.
Other approaches can be used to generate the random fields 
on non-periodic meshes and on multi-level meshes, 
see~\cite{Atzberger2010, Atzberger2010a}.

\subsection{Regime II}
\label{sec_gen_stoch_fields_regime_II}
The covariance structure can be found using 
an approach similar to the one presented
in Section~\ref{sec_gen_stoch_fields_regime_I}.
This gives
\begin{eqnarray}
\label{equ_G_regime_II}
G = 
(2k_B{T})
\left[
\begin{array}{ll}
-\rho^{-2} \Delta{x}^{-3} \OpFldDisspDiscr &               0      \\
0                                          & \OpFSDisspDiscr^{-1} \\
\end{array}
\right].
\end{eqnarray}
By factoring the covariance matrix in the Fourier basis, 
the field can be generated using FFTs in at most
$O(N\log(N))$ computational steps.

\subsection{Regime III}
\label{sec_gen_stoch_fields_regime_III}
The covariance structure can be found using 
an approach similar to the one presented
in Section~\ref{sec_gen_stoch_fields_regime_I}.
This gives 
\begin{eqnarray}
\label{equ_G_regime_III}
G = 
(2k_B{T})
\left[
\begin{array}{ll}
-\rho^{-2} \Delta{x}^{-3} \OpFldDisspDiscr & 0 \\
0                                          & 0 \\
\end{array}
\right]
.
\end{eqnarray}
By factoring the covariance matrix in the Fourier
basis, the field can be generated using FFT in 
at most 
$O(N\log(N))$ computational steps.

\subsection{Regime IV}
\label{sec_gen_stoch_fields_regime_IV}

This regime differs from the others since
the fluid momentum and structure momentum 
are both no longer represented explicitly.
Spontaneous changes in the momentum of the 
system were the primary source of 
fluctuations in the configuration of the 
structures in the other regimes.  
While the momentum is no longer represented
explicitly, we can none-the-less use a 
discretization of the momentum equations
to generate efficiently the random fields 
required in the over-damped dynamics.  
This is done by expressing the covariance of the 
stochastic driving field as
\begin{eqnarray}
\label{equ_G_regime_IV}
G = (2k_B{T})H_\subtxt{SELM} = (2k_B{T}) \left( \OpFSCoupleDiscr  \wp (-\OpFldDisspDiscr)^{-1} \wp^{\RegAdjoint} \OpFSCoupleDiscr^{\RegAdjoint} \right).
\end{eqnarray}
This makes use of $\OpSFCoupleDiscr = \OpFSCoupleDiscr^{\RegAdjoint}$ and properties of the 
specific discretized operators $\OpFldDisspDiscr$ and $\wp$.
In particular, commutativity $\wp \OpFldDisspDiscr = \OpFldDisspDiscr \wp$ and the projection 
operator properties $\wp^2 = \wp$, $\wp = \wp^{\RegAdjoint}$.  Let $U$ be a factor so that 
$UU^{\RegAdjoint} = -\OpFldDisspDiscr^{-1}$.  Using this factor we can express the covariance as 
\begin{eqnarray}
\label{equ_G_spec_factor_IV}
G = \left( \sqrt{2k_B{T}} \OpFSCoupleDiscr\wp U \right) \left( \sqrt{2k_B{T}} \OpFSCoupleDiscr\wp U \right)^{\RegAdjoint}.
\end{eqnarray}
From this expression a matrix square-root of $G$ is readily obtained,
$Q = \sqrt{2k_B{T}} \OpFSCoupleDiscr \wp U$.

We remark this is different than 
the Cholesky factor obtained for $G$ which is 
required to be lower triangular~\cite{Trefethen1997, Strang1988}.
Obtaining such a factor by Cholesky factorization would cost 
$O(M^3)$, where $M$ is the number of structure degrees of freedom.  
For the current discretization considered, the operators $L$ and 
$\wp$ are diagonalizable in Fourier space.  This 
has the consequence that the action of the operators $U$ and $\wp$ 
can be computed using FFTs with a cost of $O(N \log(N))$.  The $N$ is the 
number of lattice sites used to discretize $L$.  The stochastic driving field 
is computed from $\mathbf{h} = Q\boldsymbol{\xi}$.  This allows for the 
stochastic driving field to be generated in $O(N\log(N) + M)$ computational 
steps, assuming the action $\OpSFCoupleDiscr$ can be compute in $O(M)$ steps.  
This is in contrast to using the often non-sparse matrix arising from 
Cholesky factorization which generates the stochastic field with a cost
of $O(M^2)$.  We remark that this approach shares some similarities with the method 
proposed in~\cite{Banchio2003,Saintillan2005}.  Other methods based on splittings or 
multigrid can also be utilized to efficiently generate stochastic fields
with this required covariance structure, 
see~\cite{Atzberger2010,Atzberger2010a}.

\section{Equilibrium Statistical Mechanics of SELM Dynamics} 
\label{sec_equil_stat_mech}

We now discuss how the SELM formalism and the presented 
numerical methods capture the equilibrium statistical 
mechanics of the fluid-structure system.  This is done
through an analysis of the invariant probability distribution 
of the stochastic dynamics.  For the fluid-structure systems 
considered, the appropriate probability distribution is given 
by the Gibbs-Boltzmann distribution
\begin{eqnarray}
\label{equ_GB_distr}
\Psi_\subtxt{GB}(\mathbf{z}) = \frac{1}{Z}\exp\left[{-E(\mathbf{z})/k_B{T}}\right].
\end{eqnarray}
The $\mathbf{z}$ is the state of the system,
$E$ is the energy, $k_B$ is Boltzmann's constant, $T$
is the system temperature, and $Z$ is a normalization constant for the 
distribution~\cite{Reichl1998}.  We show this Gibbs-Boltzmann distribution 
is the equilibrium distribution of both the full stochastic 
dynamics and the reduced stochastic dynamics in each physical 
regime.

We present here both a verification of the 
invariance of the Gibbs-Boltzmann distribution for the general formalism and 
for numerical discretizations of the formalism.  
The verification is rather formal for the undiscretized formalism given technical issues
which would need to be addressed for such an infinite dimensional dynamical system.  
However, the verification is rigorous for the semi-discretization of the formalism, 
which yields a finite dimensional dynamical system.  The latter is likely 
the most relevant case in practice.  Given the nearly identical calculations involved 
in the verification for the general formalism and its semi-discretizations, we use a 
notation in which the key differences between the two cases primarily arise in the 
definition of the energy.  In particular, 
the energy is understood to be given by equation~\ref{equ_energy} 
when considering the general SELM formalism and equation~\ref{equ_energy_discr}
when considering semi-discretizations.

\subsection{Regime I}
\label{sec_equil_stat_mech_regime_I}
The stochastic dynamics given by equations~\ref{equ_SELM_I_1}--~\ref{equ_SELM_I_3} 
is a change-of-variable of the full stochastic dynamics of the SELM formalism 
given by equations~\ref{equ_SELM_0_1}--~\ref{equ_SELM_0_3}.  Thus verifying the 
invariance using the reformulated description is also applicable to equations 
~\ref{equ_SELM_0_1}--~\ref{equ_SELM_0_3} and vice versa.  To verify the 
invariance in the other regimes, it is convenient to work with the reformulated
description given for Regime I.  The energy associated with the reformulated description 
is given by
\begin{eqnarray}
E[\mathbf{p},\mathbf{v},\mathbf{X}] = \frac{1}{2\rho}\int_{\Omega} |\mb{p}(\mb{y}) - \OpSFCouple[m\mathbf{v}](\mb{y})|^2 d\mb{y} 
+ \frac{m}{2} |\mb{v}|^2 
+ \Eng[\mb{X}].
\end{eqnarray}
The energy associated with the semi-discretization is
\begin{eqnarray}
E[\mathbf{p},\mathbf{v},\mathbf{X}] = 
\frac{1}{2\rho} \sum_{\mb{m}} |\mb{p}(\mb{x}_{\mb{m}}) - \OpSFCouple[m\mathbf{v}]_{\mb{m}}|^2 \Delta\mb{x}_{\mb{m}}^3 
+ \frac{m}{2} |\mb{v}|^2 + \Eng[\mb{X}].
\end{eqnarray}

The probability density $\Psi(\mathbf{p},\mathbf{v},\mathbf{X},t)$ 
for the current state of the system under the SELM dynamics  
is governed by the Fokker-Planck equation
\begin{eqnarray}
\label{equ_Fokker_Planck_full}
\frac{\partial \Psi}{\partial t} = -\nabla \cdot \mathbf{J}
\end{eqnarray}
with probability flux
\begin{eqnarray}
\mathbf{J} & = &
\left[
\begin{array}{l}
\OpFldDissp  + \OpSFCouple + \nabla_{\mathbf{X}} \OpSFCouple\cdot \mathbf{v} + \lambda \\
-\OpFSDissp - \nabla_{\mb{X}}\Phi + \zeta\\
\mathbf{v}
\end{array}
\right]\Psi 
-\frac{1}{2}(\nabla\cdot G)\Psi
-\frac{1}{2} G \nabla \Psi.
\end{eqnarray}
The covariance operator $G$ is associated with the Gaussian field 
$\mathbf{g} = \left[\mathbf{g}_\subtxt{thm},\mathbf{F}_\subtxt{thm},0\right]^{\RegAdjoint}$
by $\langle\mathbf{g}(s)\mathbf{g}^{\RegAdjoint}(t)\rangle = G\delta(t - s)$.  In this 
regime, $G$ is given by equation~\ref{equ_SELM_I_thermal1} or~\ref{equ_G_regime_I}.  
In the notation 
$[\nabla\cdot G(\mathbf{z})]_i = \partial_{z_j} G_{ij}(\mathbf{z})$ with the summation
convention for repeated indices. To simplify the notation we 
have suppressed denoting the specific functions on which each of the operators 
act, see equations~\ref{equ_SELM_I_1}--~\ref{equ_SELM_I_3} for these details.

The requirement that the Gibbs-Boltzmann distribution $\Psi_\subtxt{GB}$ 
given by equation~\ref{equ_GB_distr} be invariant under the stochastic dynamics is equivalent 
to the distribution yielding $\nabla \cdot \mathbf{J} = 0$.  We find it convenient 
to group terms and express this condition as
\begin{eqnarray}
\label{equ_appendix_div_J}
\nabla \cdot \mathbf{J} & = & A_1 + A_2 + \nabla\cdot \mathbf{A}_3 + \nabla\cdot \mathbf{A}_4 = 0
\end{eqnarray}
where
\begin{eqnarray}
\label{equ_appendix_A_all}
\\
\nonumber
\label{equ_appendix_A_1}
A_1            & = & \left[(\OpSFCouple + 
\nabla_{\mathbf{X}} \OpSFCouple\cdot \mathbf{v} + \lambda_1)\cdot\nabla_{\mathbf{p}}E +
(-\nabla_{\mb{X}}\Phi + \zeta_1)\cdot\nabla_{\mathbf{v}}E +
(\mb{v})\cdot\nabla_{\mathbf{X}}E
\right](-k_B{T})^{-1}\Psi_\subtxt{GB} \\
\label{equ_appendix_A_2}
\nonumber
A_2            & = & \left[\nabla_{\mathbf{p}}\cdot(\OpSFCouple  + \nabla_{\mathbf{X}} \OpSFCouple\cdot \mathbf{v} + \lambda_1) 
+ \nabla_{\mathbf{v}}\cdot(-\nabla_{\mb{X}} \Eng + \zeta_2)
+ \nabla_{\mathbf{X}}\cdot(\mb{v})
\right]\Psi_\subtxt{GB} \\
\label{equ_appendix_A_3}
\nonumber
\mathbf{A}_3   & = & -\frac{1}{2}(\nabla\cdot G)\Psi_\subtxt{GB} \\
\label{equ_appendix_A_4}
\nonumber
\mathbf{A}_4   & = & \left[
\begin{array}{l}
\mathcal{L}\mathbf{u} + \lambda_2 +\left[{G_{\mb{p}\mb{p}}\nabla_{\mathbf{p}}E +
G_{\mb{p}\mb{v}}\nabla_{\mathbf{v}}E + G_{\mb{p}\mb{X}}\nabla_{\mathbf{X}}E}\right]({2k_B{T}})^{-1}
\\
-\OpFSDissp + \zeta_2 +
\left[{G_{\mb{v}\mb{p}}\nabla_{\mathbf{p}}E +
G_{\mb{v}\mb{v}}\nabla_{\mathbf{v}}E + G_{\mb{v}\mb{X}}\nabla_{\mathbf{X}}E}\right]({2k_B{T}})^{-1} \\
\left[{G_{\mb{X}\mb{p}}\nabla_{\mathbf{p}}E +
G_{\mb{X}\mb{v}}\nabla_{\mathbf{v}}E + G_{\mb{X}\mb{X}}\nabla_{\mathbf{X}}E}\right]({2k_B{T}})^{-1}
\end{array}
\right]\Psi_\subtxt{GB}.
\end{eqnarray}
We assume here that the Lagrange multipliers can be split 
$\lambda = \lambda_1 + \lambda_2$ and $\zeta = \zeta_1 + \zeta_2$
to impose the constraints by considering in isolation different terms 
contributing to the dynamics, see equation~\ref{equ_appendix_A_all}.  
This is always possible for linear constraints.  The block entries of the covariance operator $G$ are denoted
by $G_{i,j}$ with $i,j \in \{\mathbf{p},\mathbf{v},\mathbf{X}\}$.
For the energy of the discretized system given by equation~\ref{equ_energy} we have 
\begin{eqnarray}
\label{equ_appendix_grad_p_E}
\nabla_{\mathbf{p}_{\mb{n}}}E & = & \mb{u}(\mb{x}_{\mb{n}})
\Delta{x}_{\mb{n}}^3
\\
\label{equ_appendix_grad_v_E}
\nabla_{\mb{v}_q}E & = & 
\sum_{\mb{m}} 
\mb{u}(\mb{x}_{\mb{m}})
\cdot
\left(-\nabla_{\mb{v}_q}\OpSFCouple[m\mb{v}]_{\mb{m}} \right)
\Delta{x}_{\mb{m}}^3
+ m\mathbf{v}_q \\
\label{equ_appendix_grad_X_E}
\nabla_{\mb{X}_q}E & = & 
\sum_{\mb{m}} 
\mb{u}(\mb{x}_{\mb{m}})
\cdot
\left(-\nabla_{\mb{X}_q}\OpSFCouple[m\mb{v}]_{\mb{m}} \right)
\Delta{x}_{\mb{m}}^3
+ \nabla_{\mathbf{X}_q}\Eng.
\end{eqnarray}
where $\mathbf{u} = \rho^{-1}(\mathbf{p} - \OpSFCouple[m\mathbf{v}])$.
Similar expressions for the energy of the undiscretized formalism 
can be obtained by using the calculus of variations~\cite{Gelfand2000}.

We now consider $\nabla \cdot \mathbf{J}$ and each term $A_1, A_2, 
\mb{A}_3, \mb{A}_4$.
The term $A_1$ can be shown to be the time derivative of the energy 
$A_1 = dE/dt$ when considering only a subset of the contributions
to the dynamics.  Thus, conservation of the energy 
under this restricted dynamics would result in $A_1$ being 
zero. For the SELM formalism, we find by direct substitution of the gradients of $E$ given 
by equations~\ref{equ_appendix_grad_p_E}--~\ref{equ_appendix_grad_X_E}
into equation~\ref{equ_appendix_A_all} that $A_1 = 0$.  When there are 
constraints, it is important to consider only admissible states 
$(\mathbf{p}, \mathbf{v}, \mathbf{X})$.  This shows 
in the inviscid and zero temperature limit of SELM
the resulting dynamics are non-dissipative.  This property imposes constraints
on the coupling operators and can be viewed as a further motivation for the 
adjoint conditions imposed in equation~\ref{equ_adjoint_cond}.

The term $A_2$ gives the compressibility of the phase-space 
flow generated by the non-dissipative dynamics of the SELM
formalism.  The flow is generated by the vector field 
$(\OpSFCouple + \nabla_{\mb{X}}\OpSFCouple\cdot\mb{v} + \lambda_1,\hspace{0.1cm}-\nabla_{\mb{X}}\Phi + \zeta_1, \hspace{0.1cm}\mb{v})$ 
on the phase-space $(\mathbf{p}, \mathbf{v}, \mathbf{X})$.  When this
term is non-zero there are important implications for the Liouville Theorem
and statistical mechanics of the system~\cite{Tuckerman1999}.  For
the current regime, we have $A_2 = 0$ since in the divergence 
each component of the vector field is seen to be independent 
of the variable on which the derivative is computed.
This shows in the inviscid and zero temperature limit
of SELM, the phase-space flow is incompressible.  For the 
reduced SELM descriptions, we shall see this is not always the case.

The term $\mb{A}_3$ corresponds to fluxes arising from 
multiplicative features of the stochastic driving fields.  
When the covariance $G$ has a dependence on the current state 
of the system, this can result in possible changes in the amplitude 
and correlations in the fluctuations.  These changes can 
yield asymmetries in the stochastic dynamics which manifest 
as a net probability flux.  In the 
SELM formalism it is found that in the divergence of $G$ each 
contributing entry is independent of the variable on which the 
derivative is being computed.  This shows for the SELM 
dynamics there is no such probability fluxes, $\mb{A}_3 = 0$.

The last term $\mb{A}_4$ accounts for the fluxes arising from
the primarily dissipative dynamics and the stochastic driving fields.
This term is calculated by substituting the gradients of the energy
given by equation~\ref{equ_appendix_grad_p_E}--~\ref{equ_appendix_grad_X_E}
and using the choice of covariance structure given by 
equations~\ref{equ_SELM_I_thermal1} or~\ref{equ_G_regime_I}.  By
direct substitution this term is found to be zero, $\mb{A}_4 = 0$.

This shows the invariance of the Gibbs-Boltzmann distribution under 
the SELM dynamics.  This provides a rather strong validation of the 
stochastic driving fields introduced for the SELM formalism.  This shows the 
SELM stochastic dynamics are consist with equilibrium statistical 
mechanics~\cite{Reichl1998}.


\subsection{Regime II}
\label{sec_equil_stat_mech_regime_II}
For the reduced stochastic dynamics given by equations~\ref{equ_SELM_I_1}--~\ref{equ_SELM_I_3},
the probability density $\Psi(\mathbf{p},\mathbf{X},t)$ satisfies the
Fokker-Planck equation with the probability flux 
\begin{eqnarray}
\label{equ_flux_II}
\mathbf{J} & = &
\left[ 
\begin{array}{l}
\rho^{-1}\OpFldDissp  + \OpSFCouple + (\nabla_{\mb{X}}\cdot\OpSFCouple) k_B{T} + \lambda \\
\rho^{-1}\OpFSCouple + \OpFSDissp^{-1} + \zeta
\end{array}
\right]\Psi_\subtxt{GB} - \frac{1}{2}(\nabla \cdot G)\Psi_\subtxt{GB} - \frac{1}{2} G \nabla \Psi_\subtxt{GB}.
\end{eqnarray}
The $G$ denotes the covariance operator for the stochastic driving fields given by equation~\ref{equ_G_regime_II}.
The invariance of the Gibbs-Boltzmann distribution requires 
\begin{eqnarray}
\\
\nonumber
\label{equ_appendix_div_J2}
\nabla \cdot \mathbf{J} & = & A_1 + A_2 + \nabla\cdot \mathbf{A}_3 + \nabla\cdot \mathbf{A}_4 = 0 \\
\nonumber
A_1            & = & \left[(\OpSFCouple + (\nabla_{\mb{X}}\cdot \OpSFCouple)k_B{T} + \lambda_1)\cdot\nabla_{\mathbf{p}}E 
+ (\rho^{-1}\OpFSCouple + \zeta_1)\cdot\nabla_{\mathbf{X}}E
\right](-k_B{T})^{-1}\Psi_\subtxt{GB} \\
\nonumber
A_2            & = & \left[\nabla_{\mathbf{p}}\cdot(\OpSFCouple + (\nabla_{\mb{X}}\cdot \OpSFCouple)k_B{T} + \lambda_1) 
+ \nabla_{\mathbf{X}}\cdot(\rho^{-1}\OpFSCouple + \zeta_1)\right]\Psi_\subtxt{GB} \\
\nonumber
\mathbf{A}_3   & = & -\frac{1}{2}(\nabla \cdot G)\Psi_\subtxt{GB} \\
\nonumber
\mathbf{A}_4   & = & \left[
\begin{array}{l}
(\rho^{-1}\mathcal{L} + \lambda_2) +\left[{G_{\mb{p}\mb{p}}\nabla_{\mathbf{p}}E +
G_{\mb{p}\mb{X}}\nabla_{\mathbf{X}}E}\right]({2k_B{T}})^{-1}
\\
(\OpFSDissp^{-1} + \zeta_2) +
\left[
G_{\mb{X}\mb{p}}\nabla_{\mathbf{p}}E + G_{\mb{X}\mb{X}}\nabla_{\mathbf{X}}E\right]({2k_B{T}})^{-1} \\
\end{array}
\right]\Psi_\subtxt{GB}.
\end{eqnarray}
To simplify the notation we have suppressed explicitly denoting the functions on which the 
operators act, which can be inferred from equation~\ref{equ_SELM_II_1}--~\ref{equ_SELM_II_2}.  
In the current regime $m = 0$ and the energy given by equation~\ref{equ_energy_discr} has gradients 
given by
\begin{eqnarray}
\label{equ_appendix_grad_p_E2}
\nabla_{\mb{p}_{\mb{n}}}E & = & \mathbf{u}_\mb{n}\Delta{x}_{\mb{n}}^3 \\
\label{equ_appendix_grad_X_E2}
\nabla_{\mb{X}_q}E & = & \nabla_{\mb{X}_q}\Eng.
\end{eqnarray}
Similar expressions can be obtained for the undiscretized formalism using the calculus of variations~\cite{Gelfand2000}.

We now consider $\nabla \cdot \mb{J}$ and $A_1, A_2, \mb{A}_3, \mb{A}_4$.  The terms have a similar
interpretation as in Section~\ref{sec_equil_stat_mech_regime_I}.  The $A_1$ term can be interpreted
as the time derivative of the energy $A_1 = dE/dt$ when considering only a subset of the 
contributions to the dynamics.  By direct substitution of the gradients given by 
equation~\ref{equ_appendix_grad_p_E2}--~\ref{equ_appendix_grad_X_E2},
we find $A_1 = -((\nabla_{\mb{X}} \cdot \OpSFCouple)\cdot\nabla_{\mb{p}}E) \Psi_\subtxt{GB}$.  
This differs from the non-reduced equations in which this term was zero, see 
Section~\ref{sec_equil_stat_mech_regime_I}.

The term $A_2$ gives the 
compressibility of the flow generated by the vector field 
$(\OpSFCouple + (\nabla_{\mb{X}}\cdot \OpSFCouple)k_B{T} + \lambda_1, \rho^{-1}\OpFSCouple + \zeta_1)$
on the phase-space $(\mathbf{p},\mathbf{X})$.  For the 
reduced equations, the phase-space flow has compressibility given by 
$A_2 = (\rho^{-1}\nabla_{\mb{X}}\cdot \OpFSCouple)\Psi_\subtxt{GB}$,
which in general is no longer zero.  However, we have that $A_1 + A_2 = 0$.  
This follows from the form of the gradients given by 
equation~\ref{equ_appendix_grad_p_E2}--~\ref{equ_appendix_grad_X_E2}
and from the properties of $\OpFSCouple$ and $\OpSFCouple$.  
In particular, that the operators are linear and that they 
are adjoints $\OpFSCouple = \OpSFCouple^{\OpAdjoint}$ in the 
sense of equation~\ref{equ_adjoint_cond}.

The term $\mb{A}_3$ accounts for probability fluxes driven by
multiplicative features of the stochastic driving fields.  
It is found this term is zero $\mb{A}_3 = 0$.  This follows
from the divergence in which each entry of $G$ is 
independent of the variable on which the derivative is applied.  
The term $\mb{A}_4$ accounts for fluxes arising from 
the dissipative contributions to the dynamics and
the stochastic driving fields.  By direct substitution
of the gradients given in equation~\ref{equ_appendix_grad_p_E2}--~\ref{equ_appendix_grad_X_E2}, 
and the choice made for $G$ given in equation~\ref{equ_G_regime_II}, we 
find this term is zero, $\mb{A}_4 = 0$.  This establishes 
$\nabla\cdot\mb{J} = 0$ and that the Gibbs-Boltzmann distribution 
is invariant for the SELM dynamics.

\subsection{Regime III}
\label{sec_equil_stat_mech_regime_III}

We now discuss briefly the reduced stochastic dynamics given 
by equations~\ref{equ_SELM_III_1}--~\ref{equ_SELM_III_3}, 
in which $\Upsilon \rightarrow \infty$.  In this regime, 
the probability flux is almost identical to equation~\ref{equ_flux_II}, 
with any terms involving $\Upsilon^{-1}$ set to zero.  With this 
substitution, it immediately follows that the 
Gibbs-Boltzmann distribution is invariant under the SELM dynamics.

\subsection{Regime IV}
\label{sec_equil_stat_mech_regime_IV}

In the over-damped regime in which the fluid
is no longer explicitly represented, we have the reduced 
stochastic dynamics given by 
equations~\ref{equ_SELM_IV_1}--~\ref{equ_SELM_IV_3}.  
The probability density $\Psi(\mathbf{X},t)$
for the current state of the system is governed 
by the Fokker-Planck equation with the probability flux 
\begin{eqnarray}
\label{equ_flux_IV}
\\
\nonumber
\mathbf{J} & = &
\left[ 
\begin{array}{l}
H_\subtxt{SELM}[-\nabla_{\mb{X}}\Phi] + {k_B{T}}(\nabla_{\mb{X}} \cdot H_\subtxt{SELM})
\end{array}
\right]\Psi_\subtxt{GB} - \frac{1}{2}(\nabla_{\mb{X}} \cdot G)\Psi_\subtxt{GB} - \frac{1}{2} G \nabla_{\mb{X}} \Psi_\subtxt{GB}.
\end{eqnarray}
In this regime, the covariance is given by $G = 2k_B{T} H_\subtxt{SELM}$, see Section~\ref{sec_derivation_regime_IV}. 
This gives $\nabla_{\mb{X}} \cdot G = 2k_B{T} \nabla_{\mb{X}} \cdot H_\subtxt{SELM}$ and 
$\frac{1}{2}G\nabla_{\mb{X}}\Psi_\subtxt{GB} = H_\subtxt{SELM}[-\nabla_{\mb{X}} \Phi] \Psi_\subtxt{GB}$.  Substituting these
expressions in equation~\ref{equ_flux_IV}, we find $\mathbf{J} = 0$.  This establishes for the over-damped regime of the SELM formalism 
that the Gibbs-Boltzmann distribution is invariant and satisfies detailed balance. 

\subsection{Summary}
For the SELM formalism, we have demonstrated in each regime that 
the Gibbs-Boltzmann distribution is invariant.  This shows the 
SELM formalism yields appropriate behaviors with respect to 
equilibrium statistical mechanics.

\section{Applications}
\label{sec_applications}

To demonstrate how the SELM formalism can be used in practice, we consider 
spherical particles which have translation and rotational degrees of freedom.  
We give specific operators representing 
the coupling of the particles and fluid.  We compare this SELM formalism with 
classical results from fluid 
mechanics.  We should mention that similar approaches for the SELM formalism 
can be applied much more generally to represent spatially extended structures,
such as filaments, membranes, or even deformable bodies.  The development 
of representations and specific coupling operators for these structures will be the 
focus of future work.
  
\subsection{Particles with Rotational and Translational Degrees of Freedom} 
To describe particles which can exhibit 
translational and rotational motions, we use the degrees of freedom $\Xcm$ 
for the center of mass and $\Ang$ for the rotational configuration.  To describe the full 
configuration of a particle, we define the composite vector $\mathbf{X} = (\Xcm,\Ang)$.
To investigate how the coupling operators capture the hydrodynamics of the system,
it is convenient to characterize the system in Regimes III and IV of 
Sections~\ref{sec_summary_SELM_regimeIII} and~\ref{sec_summary_SELM_regimeIV}.
This highlights central features of the coupling operators also relevant in
the other regimes.  Given the specific degrees of freedom of the particles, it 
is convenient to express 
equation~\ref{equ_SELM_III_2} as
\begin{eqnarray}
\frac{\partial \Xcm}{\partial t} & = & \OpFSCouple_0 \mathbf{u} \\ 
\frac{\partial \Ang}{\partial t} & = & \OpFSCouple_1 \mathbf{u}.
\end{eqnarray}
The $\mb{u} = \rho^{-1}\mb{p}$.
To represent the kinematics of such particles for a given state of the flow field 
of the fluid we use
\begin{eqnarray}
\OpFSCoupleDiscr_0 \mathbf{u} & = &
\sum_{\mathbf{m}}
\big\langle\hspace{0.1cm}
\eta_0(\mathbf{y}_{\mathbf{m}} - (\Xcm + \mathbf{z}) ) 
\hspace{0.1cm}
\mathbf{u}_{\mathbf{m}}
\hspace{0.1cm}
\big\rangle_{\tilde{\mathcal{S}}, |\mathbf{z}| = R}
\Delta{x}_{\mathbf{m}}^3 \\ 
\OpFSCoupleDiscr_1 \mathbf{u} & = & \frac{3}{2 R^2} 
\sum_{\mathbf{m}}
\big\langle\hspace{0.1cm}
\eta_1(\mathbf{y}_{\mathbf{m}} - (\Xcm + \mathbf{z}) ) \left(\mathbf{z} 
\times \mathbf{u}_{\mathbf{m}}\right) 
\hspace{0.1cm}
\big\rangle_{\tilde{\mathcal{S}}, |\mathbf{z}| = R}
\Delta{x}_{\mathbf{m}}^3.
\end{eqnarray}
The angle brackets denote an average over the surface of the sphere
which is given by the quadrature
\begin{eqnarray}
\big\langle
\hspace{0.1cm}
f(\mathbf{z})
\hspace{0.1cm}
\big\rangle_{\tilde{\mathcal{S}}, |\mathbf{z}| = R} 
& = & 
\frac{1}{4\pi R^2} \sum_{k} w_k f(\mathbf{z}_k).
\end{eqnarray}
The $w_k$ denote the quadrature weights and 
the $\mathbf{z}_k$ denote the quadrature nodes.
In practice, we use the Lebedev quadratures~\cite{Lebedev1999}.

These SELM kinematics are in fact closely related to the exact kinematics 
of a passive spherical particle expressed using the Faxen Theorem of 
fluid dynamics~\cite{Bedeaux1974}.  The expression from the Faxen Theorem
corresponds to the continuum limit of the above expressions and when
the kernel functions are replaced by Dirac 
$\delta$-functions~\cite{Bedeaux1974,Mazur1974,Hills1975}.
  
For particles which actively exert forces on the fluid, we develop 
a coupling operator for the force by using the adjoint condition 
given by equation~\ref{equ_adjoint_cond}.  In fact, this condition 
can be interpreted as requiring the fluid-structure 
coupling conserve the energy of the system in Regime III in the 
inviscid and zero temperature limit.  This is seen 
formally by letting
$\OpFldDissp \rightarrow 0$ and $T \rightarrow 0$ in equation~\ref{equ_SELM_III_1}
and computing what is required for $dE/dt = 0$.  
Using this condition, we obtain the fluid coupling operator by considering 
directly the adjoint condition $\OpSFCoupleDiscr = \OpFSCoupleDiscr^{\OpAdjoint}$.
For the specific coupling operators considered for the particles,
the adjoint condition of the discretized system gives
\begin{eqnarray}
\OpSFCoupleDiscr_0(\mathbf{x}_{\mathbf{m}}) & = & 
\left(
\big\langle\hspace{0.1cm}
\eta_0(\mathbf{x}_{\mathbf{m}} - (\Xcm + \mathbf{z})) 
\hspace{0.1cm}
\big\rangle_{\tilde{\mathcal{S}},|\mathbf{z}| = R}
\right)
\mathbf{F} \\
\OpSFCoupleDiscr_1(\mathbf{x}_{\mathbf{m}}) & = & -\frac{3}{2 R^2} 
\left(
\big\langle\hspace{0.1cm}
\mathbf{z} \eta_1(\mathbf{x}_{\mathbf{m}} - (\Xcm + \mathbf{z})) 
\hspace{0.1cm}
\big\rangle_{\tilde{\mathcal{S}},|\mathbf{z}| = R}
\right)
\times \mathbf{T}.
\end{eqnarray}
The $\mathbf{F} = -{\partial \Eng}/{\partial \mathbf{\Xcm}}$
is the total force acting on the particle and 
$\mathbf{T} = -{\partial \Eng}/{\partial \mathbf{\Ang}}$ is the 
total torque acting on the particle.
The adjoint condition holds exactly for the discretized system 
provided the same quadrature is used for both $\OpFSCoupleDiscr$
and $\OpSFCoupleDiscr$.
  
\begin{figure}[t*]
\centering
\includegraphics[width=5in]{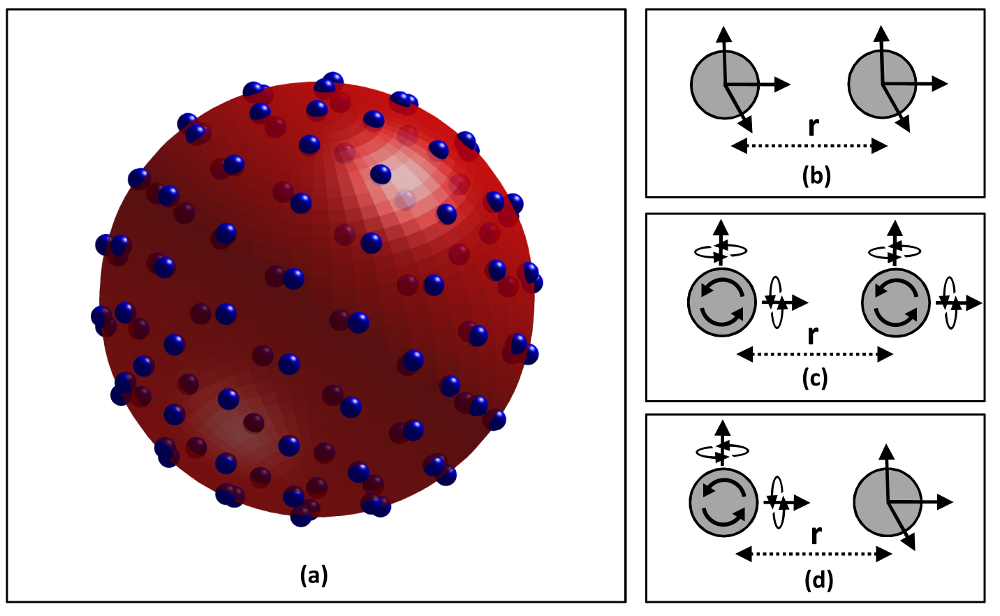}
\caption[SELM Rotational Coupling Modes]
{Translational and Rotational Motions of Spherical Particles.
On the left is shown the Lebedev quadrature nodes for $N = 110$ which
are used for computing averages on the surface of a sphere used in the 
SELM coupling operators.  On the right is shown the three
modalities of coupling for spherical particles.  These are 
Translation-Translation (b), 
Rotation-Rotation (c), and Rotation-Translation (d).
}
\label{fig_coupling_modes}
\end{figure}

In the SELM formalism, the coupling operators 
$\OpFSCoupleDiscr$ and $\OpSFCoupleDiscr$ encapsulate the 
effective hydrodynamic coupling between the 
particles and fluid.  To characterize 
how the presented SELM operators represent
such hydrodynamic coupling in practice,
we consider the interactions between two 
spherical particles.  The coupling of 
two spherical particles 
has three modes of coupling: (i) translation-translation,
(ii) rotation-rotation, and (iii) rotation-translation,
see Figure~\ref{fig_coupling_modes}.
To characterize these coupling modes, we consider
the effective hydrodynamic coupling tensor (mobility) 
which relates applied forces and torques to consequent
motions of the particles.  An effective coupling tensor 
can be determined rather naturally for the SELM formalism
by considering Regime IV at zero temperature.  In this
regime, the effective coupling tensor arises in the dynamics 
\begin{eqnarray}
\frac{d}{dt}
\left[
\begin{array}{l} 
\Xcm \\
\Ang
\end{array}
\right] 
& = &    
\left[
\begin{array}{ll} 
\tilde{H}_\subtxt{TT} & \tilde{H}_\subtxt{TR} \\
\tilde{H}_\subtxt{RT} & \tilde{H}_\subtxt{RR} 
\end{array}
\right] 
\left[
\begin{array}{l} 
\mathbf{F} \\
\mathbf{T}
\end{array}
\right]  \\
\tilde{H}_{\ell k} & = & \OpFSCoupleDiscr_\ell (\wp \OpFldDisspDiscr)^{-1} \OpSFCoupleDiscr_k.
\end{eqnarray}
The tilde is used throughout to distinguish the tensor components 
associated with the SELM formalism from those arising in
other theories from fluid mechanics.  The subscripts $T,R$ 
indicates components related respectively to the 
translational and rotational degrees of freedom.
In practice, the components of these tensors can be easily 
computed from an implementation of SELM by applying
forces or torques which are set to $\mathbf{e}_{j}$ 
on only one particle and considering the $i^{th}$ component 
of the velocity or angular velocity of the other 
particle.

To compare the results of the proposed SELM coupling 
operators with classical results of fluid mechanics,
we consider the following Rotne-Prager-Yamakawa tensors 
(RPY tensors)~\cite{Rotne1969, Yamakawa1970, Reichert2004}
\begin{eqnarray}
H_\subtxt{TT}(\mathbf{r}) 
& = & 
\left\lbrace
\begin{array}{ll}
\frac{1}{8 \pi \mu r}
\left(1 + \frac{2a^2}{3r^2}\right)\mathcal{I} 
+ 
\frac{1}{8 \pi \mu r}
\left(1 - \frac{2a^2}{r^2}\right)
\mb{\hat{r}}\mb{\hat{r}}^T
,
&\mbox{\small for $r \geq 2a$} \\
\frac{1}{16 \pi \mu a}
\left(\frac{8}{3} - \frac{3r}{4a}\right)\mathcal{I} 
+ 
\frac{r}{64 \pi \mu a^2}
\mb{\hat{r}}\mb{\hat{r}}^T
,
&\mbox{\small for $r < 2a$} 
\end{array}
\right\}\\
\label{equ_H_RR}
H_\subtxt{RR}(\mathbf{r}) & = & \frac{1}{16\pi\mu r^3} 
\left(\mathcal{I} - {3\mb{\hat{r}}\mb{\hat{r}}^T}
\right) \\
H_\subtxt{TR}(\mathbf{r}) & = & \frac{1}{8\pi\mu r^2} \mathbf{\hat{r}} \times .
\end{eqnarray}
In this notation, $\mathbf{\hat{r}} = \mathbf{r}/r$ and 
we denote by $\mathbf{\hat{r}}\times$ the matrix which 
represents the action of the cross-product of a vector 
with $\mathbf{\hat{r}}$.  The $a$ denotes the effective
hydrodynamic radius of the particle.  The RPY
tensors capture far-field interactions and 
are expected to be accurate physically only 
when the particles are sufficiently separated. 
They are not designed to capture lubrication 
interactions or other near-field 
effects~\cite{Brady1988,Acheson1990,Rotne1969,Yamakawa1970}.

\begin{figure}[t*]
\centering
\includegraphics*[width=5.25in]{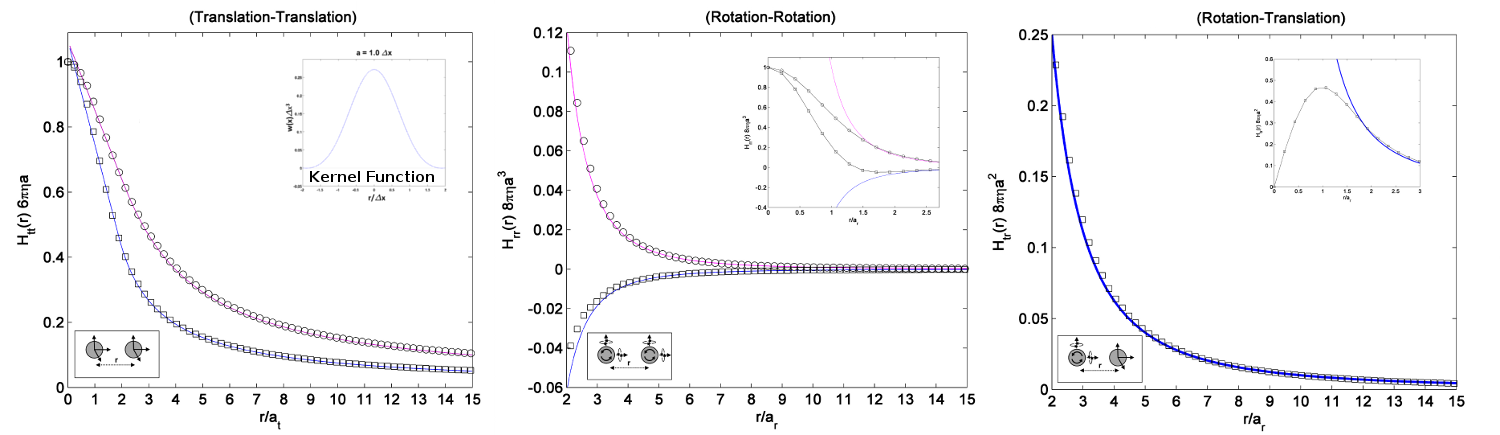}
\caption[SELM Effective Hydrodynamic Coupling Tensor]
{Effective Hydrodynamic Coupling Tensors of the SELM Operators.
The components of the effective hydrodynamic coupling tensor 
associated with the SELM operators are compared with the 
Rotne-Prager-Yamakawa Tensors of fluid mechanics. 
All components are scaled by the reference mobility 
$m_t = 1/6\pi\mu a_t$ for translation and 
$m_r = 1/8\pi\mu a_r^3$ for rotation.  The kernel functions
$\eta_0$ and $\eta_1$ were chosen to be the radial symmetric 
function shown as the inset on the left.  The other insets
show the near-field hydrodynamic interactions.
}
\label{figure_compare_SELM_H_composite}
\end{figure}

For the proposed coupling operators, very good agreement is 
found with the RPY tensors for all of the different coupling 
modes, see Figure~\ref{figure_compare_SELM_H_composite}.
This agreement is especially good for particles
separated at least a distance of two radii.
At closer distances the RPY tensors either reflect
some type of regularization or they diverge.  In the case of 
translation-translation coupling the SELM 
formalism agrees even in the near-field with the
RPY tensor~\cite{Rotne1969,Yamakawa1970}.  For the other 
cases, the RPY tensors diverge.  In all cases,
the SELM formalism provides 
as the separation distance becomes small a 
regularized tensor.  The regularization can
be interpreted as an interpolation between
the two particle far-field interaction to 
the single particle response, see insets in 
Figure~\ref{figure_compare_SELM_H_composite}.
For the rotation-translation coupling model,
we see the operators have the important property 
for the single particle response that the rotation
and translational motions are decoupled.  This 
agrees with predictions from fluid mechanics for
linear Stokes flow.  

These results for the proposed operators demonstrate 
that the SELM formalism provides a practical 
approach for simulating hydrodynamically coupled 
spherical particles.  We remark that the near-field 
artifacts are a by-product of the specific 
coupling operators utilized to approximate the 
fluid-structure interactions.  For many practical systems,
repulsive long-range interactions keep particles 
well-separated avoiding these near-field artifacts.
It should be emphasized the SELM formalism is not limited to 
such cases, since a more accurate approach which captures 
near-field effects can be developed by using a different 
choice for the coupling operators.  

While we have presented only one rather special application,
the SELM formalism can be applied more broadly.  This simply
requires appropriate representations for the structures and 
a choice for the coupling operators.  It is expected a 
wide variety of structures could be studied using this approach, 
including particles of non-spherical shape, filaments, membranes, 
and deformable bodies.  The development of coupling operators for
these systems will be the focus of future work.

\section{Conclusions}
An approach for fluid-structure interactions subject to thermal fluctuations 
was presented based on a mechanical description utilizing 
both Eulerian and Lagrangian reference frames.  General conditions were 
established for operators coupling these descriptions.  Reduced descriptions 
for the stochastic dynamics of the fluid-structure system were 
developed for several physical regimes.  Analysis was presented 
for each regime establishing for the SELM stochastic dynamics 
that the Gibbs-Boltzmann distribution is invariant.  The SELM 
formalism provides a general framework for 
the development of computational methods for applications
requiring a consistent treatment of structure mechanics, 
hydrodynamic coupling, and thermal fluctuations.

\section{Acknowledgements}
The author P.J.A. acknowledges support from research 
grant NSF CAREER DMS - 0956210.
   
\bibliographystyle{siam}
\bibliography{paperBibliography}

\appendix

\end{document}